\documentclass[journal]{IEEEtran}

\usepackage{mathrsfs}
\usepackage[noadjust]{cite}
\usepackage{graphicx,color,overpic,psfrag}
\usepackage{amsmath, amssymb}
\usepackage{cleveref}
\usepackage{float}
\usepackage{latexsym}
\usepackage{bm}
\usepackage{amssymb}
\usepackage{cases}
\usepackage{array}
\usepackage{fancyhdr}
\usepackage{setspace}

\ifCLASSOPTIONcompsoc
\usepackage[caption=false,font=normalsize,labelfont=sf,textfont=sf]{subfig}
\else
\usepackage[caption=false,font=footnotesize]{subfig}
\fi

\definecolor{myblue}{rgb}{.91,.95,.99}

\usepackage{url}
\usepackage{algorithm}
\usepackage{algorithmic}

\usepackage{blkarray}
\usepackage{booktabs}

\usepackage{multirow}
\usepackage{dsfont}
\usepackage{tabularx}
\usepackage[table]{xcolor}
\usepackage{amsfonts}

\usepackage{letltxmacro}

\graphicspath{{figure/}}

\newcommand{\ra}[1]{\renewcommand{\arraystretch}{#1}}
\newcolumntype{L}{>{\hspace*{-\tabcolsep}}l}
\newcolumntype{R}{c<{\hspace*{-\tabcolsep}}}
\definecolor{lightblue}{rgb}{0.93,0.95,1.0}

\newtheorem{lemma}{Lemma}

\newcommand{\tabref}[1]{Table \ref{#1}}
\newcommand{\alref}[1]{Algorithm \ref{#1}}
\newcommand{\appref}[1]{Appendix \ref{#1}}

\newcommand{\blkdiag}[1]{\mathsf{blkdiag}\left\{#1\right\}}

\newcommand{\abs}[1]{\left|#1\right|}

\newcommand{\thetabs}[2]{{\dnnot{\theta}{bs}}}

\newcommand{\bb}{\mathbf{b}}
\newcommand{\bc}{\mathbf{c}}

\newcommand{\bh}{\mathbf{h}}

\newcommand{\bp}{\mathbf{p}}

\newcommand{\br}{\mathbf{r}}
\newcommand{\bs}{\mathbf{s}}

\newcommand{\bu}{\mathbf{u}}
\newcommand{\bv}{\mathbf{v}}
\newcommand{\bw}{\mathbf{w}}

\newcommand{\bA}{\mathbf{A}}
\newcommand{\bB}{\mathbf{B}}
\newcommand{\bC}{\mathbf{C}}
\newcommand{\bD}{\mathbf{D}}

\newcommand{\bI}{\mathbf{I}}

\newcommand{\bK}{\mathbf{K}}

\newcommand{\bV}{\mathbf{V}}
\newcommand{\bW}{\mathbf{W}}
\newcommand{\bX}{\mathbf{X}}

\newcommand{\bbC}{\mathbb{C}}

\newcommand{\sinr}{\mathtt{SINR}}

\newcommand{\dnnot}[2]{#1_{\mathrm{#2}}}

\allowdisplaybreaks

\begin{document}

\title{Hybrid Analog/Digital Precoding for Downlink Massive MIMO LEO Satellite Communications}
\author{
Li~You,
Xiaoyu~Qiang,
Ke-Xin~Li,
Christos~G.~Tsinos,
Wenjin~Wang,
Xiqi~Gao,
and~Bj\"{o}rn~Ottersten %

\thanks{Copyright (c) 2015 IEEE. Personal use of this material is permitted. However, permission to use this material for any other purposes must be obtained from the IEEE by sending a request to pubs-permissions@ieee.org.}%
\thanks{This work was presented in part at ICC'2021 \cite{qiang2021hybrid}.}
\thanks{
Li You, Xiaoyu Qiang, Ke-Xin Li, Wenjin Wang, and Xiqi Gao are with the National Mobile Communications Research Laboratory, Southeast University, Nanjing 210096, China, and also with the Purple
Mountain Laboratories, Nanjing 211100, China (e-mail: lyou@seu.edu.cn; xyqiang@seu.edu.cn; likexin3488@seu.edu.cn; wangwj@seu.edu.cn; xqgao@seu.edu.cn).
}
\thanks{
Christos G. Tsinos and Bj\"{o}rn Ottersten are with the University of Luxembourg, Luxembourg City 2721, Luxembourg (e-mail: chtsinos@gmail.com; bjorn.ottersten@uni.lu).
}
}

\maketitle

\begin{abstract}
Massive multiple-input multiple-output (MIMO) is promising for low earth orbit (LEO) satellite communications due to the potential in enhancing the spectral efficiency. However, the conventional fully digital precoding architectures might lead to high implementation complexity and energy consumption.
In this paper, hybrid analog/digital precoding solutions are developed for the downlink operation in LEO massive MIMO satellite communications, by exploiting the slow-varying statistical channel state information (CSI) at the transmitter. First, we formulate the hybrid precoder design as an energy efficiency (EE) maximization problem by considering both the continuous and discrete phase shift networks for implementing the analog precoder. The cases of both the fully and the partially connected architectures are considered. Since the EE optimization problem is nonconvex, it is in general difficult to solve. To make the EE maximization problem tractable, we apply a closed-form tight upper bound to approximate the ergodic rate. Then, we develop an efficient algorithm to obtain the fully digital precoders. Based on which, we further develop two different efficient algorithmic solutions to compute the hybrid precoders for the fully and the partially connected architectures, respectively. Simulation results show that the proposed approaches achieve significant EE performance gains over the existing baselines, especially when the discrete phase shift network is employed for analog precoding.
\end{abstract}


\section{Introduction}\label{sec:net_intro}
Satellites are suitable for providing services for areas where a terrestrial infrastructure is nonexistent, thus playing an indispensable part in the wider adoption of future wireless networks \cite{guidotti2019architectures}. So far, an integrated satellite-terrestrial scheme has been proposed to meet the comprehensive and diverse service requirements of 5G, such as maintaining a global network \cite{jia2016broadband}. Due to the increasing demand for higher throughput and data rate services, ubiquitous coverage has been attained through the wide adoption of multi-beam satellite communication (SATCOM) systems \cite{wang2019near}. Low earth orbit (LEO) satellites, located from 600 to 1500 km, have attracted great interest due to their relatively shorter propagation delay, smaller path loss, and lower production and launching costs in contrast to geostationary earth orbit (GEO) or medium earth orbit (MEO) counterparts \cite{qu2017leo}. Recently, many LEO high-throughput satellite constellations including OneWeb, Telesat, and Starlink have been developed \cite{su2019broadband}.

Massive multiple-input multiple-output (MIMO) transmission has been widely adopted in terrestrial communications to provide high performance gains and enhanced coverage \cite{angeletti2020pragmatic}. However, it has not gained much attention so far in SATCOM systems. In this paper, we study the application of the massive MIMO technology in LEO SATCOM systems, where the LEO satellite is equipped with a large amount of antennas. With multiple available degrees of freedom in the spatial domain, a massive MIMO system can achieve significant spectral and energy efficiency (EE) gains, making it a promising technology for future satellite systems \cite{you2020massive,li2020downlink}.

In general, satellites are mainly powered by solar panels and when sun is blocked by the earth, they utilize internal batteries to operate.   Hence, the energy consumed by a SATCOM system is nonnegligible while the satellite payload has limited capabilities in its energy consumption \cite{gao2020robust}, as well. Thus, it is meaningful to design the system based on a performance metric that aims at the best performance under the least possible power consumption. The majority of the previous works focused on maximizing the transmission rate of the SATCOM system, a performance metric that is able to satisfy the first requirement, i.e., the best performance, only \cite{Fuchs2018performance, li2020downlink}. A more suitable metric that is able to effectively satisfy both requirements, is the EE one. To that end, the precoder design of the LEO SATCOM system is based on optimization of the EE metric in the present paper.

The accuracy of channel state information (CSI) that can be obtained is critical for the performance of a precoding method.
Most existing works are developed under the assumption of perfect knowledge of the instantaneous CSI (iCSI). However, in practice, accurate iCSI at the transmitter during the downlink operation of a SATCOM system is quite difficult to obtain due to several reasons. In general, both the user terminals (UTs) and the satellite are moving, leading to large Doppler shifts. Also, the propagation delay between the UTs and the satellite is long, especially compared with that in the terrestrial systems \cite{you2020massive}.
Specifically, as the downlink of LEO SATCOM systems functions in the frequency division duplexing (FDD) mode, the iCSI is estimated at the UTs and then fed back to the satellite. The corresponding overhead for pilot training and channel estimation is large and no doubt occupies considerable resources. As a result of that inefficiency, the received iCSI at the satellite terminal is out-of-date. Besides, in the time division duplexing (TDD) mode, the downlink iCSI utilized at the satellite terminal is obtained according to the reciprocity between the uplink and downlink \cite{zhang2015large}. However, this approach is applied under the assumption that the downlink and uplink channels are the same within the channel coherence time \cite{You15Pilot}, which might be less than the transmission delay, leading to nonnegligible errors. The above facts motivate us to design the precoding signals of the downlink LEO SATCOM system under consideration based on the assumption of the statistical CSI (sCSI) knowledge. Note that the sCSI can be assumed to be fixed for a relatively long time period \cite{li2020downlink}.

In conventional SATCOM systems, a fully digital precoding method, applied at the baseband, requires one radio frequency (RF) chain per antenna element, which presents high hardware complexity and is also energy-inefficient for massive MIMO. Meanwhile, although existing fully analog precoding solutions require only a single RF chain transmitter \cite{yu2016alternating}, they can only support single stream communications. Contrariwise to these extreme cases, hybrid precoding is able to support multi-stream communications with a much smaller number of RF chains than the antenna elements.

In literature so far, there are two major architectures for hybrid precoding, namely the fully and the partially connected ones, which are determined by the way the RF chains are connected to the antenna array \cite{yu2016alternating}. In the fully connected architecture, each antenna element is connected to each one of the available RF chains. The major disadvantage of the fully connected architecture is the large number of phase shifters required to achieve analog precoding \cite{gao2016energy}. This results in high power consumption that a SATCOM system might not be able to support. As for the partially connected architecture, each RF chain is only connected to a part of the antenna array. Such an approach requires much fewer phase shifters to implement the analog precoder compared to the fully connected architecture, resulting in a SATCOM transmitter of decreased hardware complexity and power consumption. Another aspect that should be taken into account is the hardware complexity of the phase shifters. In practice, phase shifters of high resolution are expensive components \cite{wang2018hybrid} leading to a SATCOM system of high cost. 

So far, hybrid precoding design on terrestrial communications has been extensively investigated based on the rate maximization criterion \cite{yu2016alternating,liu2015two,sohrabi2016hybrid,kwon2018millimeter,kwon2020limited}. Concerning EE, an alternative way to design the hybrid precoders has also been studied extensively for terrestrial communications in e.g., \cite{kaushik2020joint, kaushik2019dynamic, tsinos2017energy, zi2016energy}. For example, in \cite{zi2016energy}, the authors investigated the optimization of EE for 5G wireless communication systems with a large number of antennas and RF chains. In general, precoding over SATCOM has also been a well-studied topic, though the majority of the existing works focus on the fully digital architectures. In \cite{you2020massive}, the authors proposed a massive MIMO transmission scheme for LEO SATCOM systems with sCSI.  In \cite{wang2018robust}, the authors studied the maximization of the minimum average signal-to-interference-plus-noise ratio (SINR) under the per beam power constraints for multibeam SATCOM systems. In \cite{you2018outage}, the case of outage constrained robust multigroup multicast beamforming was investigated.
There are existing hybrid precoding works for SATCOM systems, but most of them are based on the optimization of the spectral efficiency \cite{arora19hybrid}.

In addition, a large amount of research works have been done on hybrid precoding with the fully and/or the partially connected architectures on terrestrial communications, but most of them assume that infinite resolution phase shifters are employed for the implementation of analog precoders. For example, manifold optimization and semidefinite relaxation based alternation minimization algorithms were proposed in \cite{yu2016alternating} for the two architectures, respectively. Besides, in \cite{arora2019hybrid}, the authors proposed two types of algorithms based on the majorization-minimization (MM) framework for these architectures. In \cite{he2016energy}, the authors investigated the design of a hybrid precoder and a combiner under EE maximization criterion with the partially connected architecture in terrestrial millimeter-wave communications. Research on finite resolution phase shifters for hybrid precoding has attracted interest recently in e.g., \cite{wang2018hybrid,cui2019hybrid,chen2017hybrid,chen2018low}, but they employ either exhaustive search or nearest point projection (NPP) method, leading to high computational complexity or inaccuracy.  Thanks to recent
advancements in discrete phase optimization, in \cite{mendezrial2016hybrid}, a framework has been established for designing multiuser precoding under the one-bit, continuous constant-envelope, and discrete constant-envelope cases, respectively, which runs faster and has higher accuracy compared to the previously existed works \cite{wang2018hybrid,cui2019hybrid,chen2017hybrid,chen2018low}.

Inspired by the aforementioned studies, we focus on hybrid precoding for the fully and the partially connected architectures under the EE maximization criterion for downlink massive MIMO LEO SATCOM systems.
The main contributions of this paper are summarized as follows:
\begin{itemize}
  \item We formulate the problem of energy efficient hybrid precoding for massive MIMO LEO SATCOM exploiting sCSI at the transmitter. The ergodic sum rate in the numerator of the objective function, i.e., EE, is nonconvex and involves the expectation operator, which does not admit an easy-to-tackle expression. It can be estimated through the application of a Monte Carlo method but the complexity is high. Therefore, we adopt an upper bound on the ergodic sum rate which is easier to tackle in the latter optimization problem.
  \item The EE maximization is a fractional programming problem with a nonconvex constraint considering both the continuous and discrete phase shift networks for implementing the analog precoder, and the optimization variables, i.e., analog and digital precoders, are tightly coupled, which is in general, hard to solve. Hence, we first consider the product of these two
      precoders as a whole. Then, we develop an efficient algorithmic solution based on Dinkelbach's algorithm and the weighted minimum mean-square error (WMMSE) method to tackle the corresponding fully digital precoding design problem. Finally, the analog and the digital precoders can be equivalently obtained by minimizing the Euclidean distance between the fully digital and the hybrid precoders, which is still a nonconvex problem. We tackle it by alternating optimization of the analog and the digital precoders.
  \item For the fully connected architecture, we first fix the analog precoder and then, the digital precoder can be directly derived by handling a least squares problem. Then, with the digital precoder fixed, we optimize the analog precoder. Since it is still a tricky problem considering the phase shift constraint for the analog precoder, we develop an inexact accelerated projected gradient (APG)-based MM algorithm to tackle it.
  \item For the partially connected architecture, we first derive a closed-form solution of the digital precoder through variable projection. We then obtain an equivalent formulation after substituting the expression of the digital precoder back to the objective function and vectorizing the analog precoder. The simplified nonconvex problem can be tackled through the inexact MM algorithm in a manner similar to the case of the fully connected architecture.
  \item Numerical results demonstrate that the proposed approaches achieve significant EE performance gains over the existing baselines for both the fully and the partially connected architectures, especially when discrete phase shifters are adopted for analog precoding.
\end{itemize}

The rest of this paper is organized as follows. The system model and problem formulation are presented in Section \ref{sec:sys_mod}. Section \ref{sec:fullydigital_solution} proposes an algorithm to tackle this problem. The solutions for the fully and the partially connected architectures are presented in Sections \ref{sec:fullyconnected} and \ref{sec:partiallyconnected}, respectively. Simulation results are presented in Section \ref{sec:sim}, and Section \ref{sec:conc} concludes the paper.

\emph{Notations}: Column vectors and matrices are represented by the lower and upper case boldface letters, respectively. Conjugate, Hermitian conjugate, and transpose operations are denoted by $(\cdot)^\ast$, $(\cdot)^H$, and $(\cdot)^T$, respectively. $\blkdiag{\bX_1,\ldots,\bX_N}$ denotes the block diagonal matrix whose principal diagonal blocks are $\bX_1,\ldots,\bX_N$. Kronecker product is represented by $\otimes$. The matrix trace operator is $\mathrm{Tr}\{\cdot\}$. The expectation operator is denoted by $\mathbb{E}\{\cdot\}$ and $\exp\{\cdot\}$ is the exponential operator. The $\ell_2$-norm and Frobenius-norm are denoted by $||\cdot||_2$ and $||\cdot||_F$, respectively. The magnitude, angle, real and imaginary parts of a complex number $x$ are $|x|$, $\angle x$, $\Re(x)$ and $\Im(x)$, respectively. The ceil and floor values of $x$ are denoted by $\lceil x \rceil$ and $\lfloor x \rfloor$, respectively. The unitary space with $m\times n$-dimension is represented by $\mathbb{C}^{m\times n}$ and $\jmath=\sqrt{-1}$ denotes the imaginary unit. The gradient of a function concerning complex variable $x$ is defined as $\nabla_xf(x) = \nabla_{\Re(x)}f(x) + \jmath\nabla_{\Im(x)}f(x)$. The $N$-dimensional identity matrix is denoted by $\bI_N$, where the subscript is generally omitted for brevity. The $(m, n)$th element of matrix $\bA$ is represented by $\bA_{m,n}$ and $\bA_{i:j,\ell:k}$ denotes the submatrix constituted by the $i$th to the $j$th rows and the $\ell$th to the $k$th columns of $\bA$. The real-valued and the complex-valued Gaussian distributions with mean vector ${\bm{\mu}}$ and covariance matrix $\bK$ are denoted as $\mathcal{N}(\bm{\mu},\bK)$ and $\mathcal{CN}(\bm{\mu},\bK)$, respectively. The operators $\triangleq$ and $\sim$ represent ``be defined as" and ``be distributed as", respectively. The computational complexity is denoted by $\mathcal{O}(\cdot)$.

\section{System Model and Problem Formulation}\label{sec:sys_mod}

\subsection{Channel Model}
\begin{figure}[htb]
		\centering
		\includegraphics[width=0.45\textwidth]{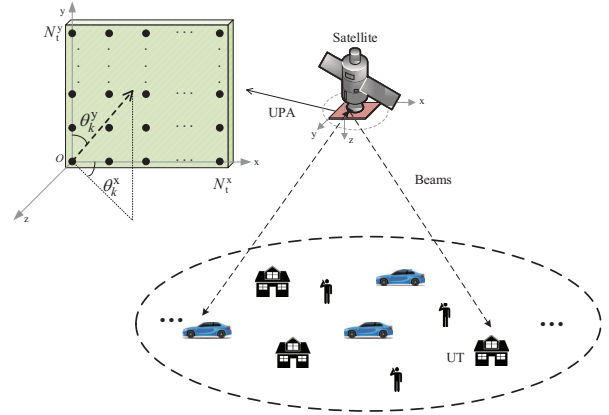}
		\caption{A downlink LEO SATCOM system.}
        \label{COM_SYS}
	\end{figure}

Consider a downlink LEO SATCOM system with $K$ single-antenna UTs, as illustrated in Fig. \ref{COM_SYS}. A large-scale uniform planar array (UPA) is applied at the satellite, constituted by $N_\mathrm{t}^{\mathrm{x}}$ and $N_\mathrm{t}^{\mathrm{y}}$ elements with half-wavelength separation on the x- and y-axes, respectively. We denote $N_\mathrm{t}\triangleq N_\mathrm{t}^{\mathrm{x}}N_\mathrm{t}^{\mathrm{y}}$ as the number of antennas at the LEO satellite.

In this work, a multi-path model is adopted to characterize the downlink channel between the satellite and the $k$th UT, with multipath number being $L_k$. Then, with a ray-tracing based approach, at time instant $t$ and frequency $f$, the complex-valued baseband channel frequency response is given by \cite{you2020massive}
\begin{align}\label{eq:orgcm}
\bc_k\left(t,f\right)=\sum_{l=1}^{L_{\rm k}}\alpha_{k,l}\exp\{\jmath 2\pi[t\nu_{k,l}-f\tau_{k,l}]\}\bv_{k,l},
\end{align}
where the complex parameters associated with path $l$ of UT $k$, $\alpha_{k,l}$, $\nu_{k,l}$, $\tau_{k,l}$, and $\bv_{k,l}$, denote the channel gain, the Doppler shift, the propagation delay, and the array response vector, respectively.
Then, we characterize these parameters based on the channel properties of the considered LEO SATCOM system.

In particular, the major component of $\nu_{k,l}$ is the Doppler shifts arising from the mobility of the satellite and the UT, which is represented by $\nu_{k,l}=\nu_{k,l}^{\rm sat}+\nu_{k,l}^{\rm ut}$.
Note that for SATCOM channels, due to the much higher altitude of the satellite compared with that of the scatterers located a few kilometers near the terrestrial UTs, the Doppler shifts of the satellite and the array response vectors of all propagation paths can be assumed to be identical, i.e., $\nu_{k,l}^{\rm sat}=\nu_{k}^{\rm sat}$ and $\bv_{k,l}=\bv_k$, $\forall k,l$ \cite{you2020massive,3gpp}.
Besides, we denote the propagation delay as $\tau_{k,l}=\tau_{k,l}^{\rm ut}+\tau_k^{\rm min}$, where $\tau_k^{\rm min}=\min_{l}\tau_{k,l}$ is the minimum delay among all the prorogation paths of the $k$th UT.
Then, the channel response in \eqref{eq:orgcm} can be converted into
\begin{align}\label{eq:orgcmr}
\bc_k\left(t,f\right)=\exp\{\jmath 2\pi[t\nu_k^{\rm sat}-f\tau_k^{\rm min}]\}g_k(t,f)\bv_k.
\end{align}
The UPA response vector $\bv_k$ can be expressed as
    \begin{equation}
    \begin{aligned}\label{eq:arvk}
    \mathbf{v}_k=\mathbf{v}_k^{\mathrm{x}}\otimes \mathbf{v}_k^{\mathrm{y}}=\mathbf{v}_{\mathrm{x}}(\vartheta_k^{\mathrm{x}})\otimes\mathbf{v}_{\mathrm{y}}(\vartheta_k^{\mathrm{y}})\in \mathbb{C}^{N_t\times 1},
    \end{aligned}
    \end{equation}
    where $\mathbf{v}_k^d$ for $d\in \mathcal{D}\triangleq \{x,y\}$ is the array response vector of the angle with respect
to the x- or y-axis given by
    \begin{equation}
    \begin{aligned}\label{eq:crvc}
\mathbf{v}_k^d &\triangleq \mathbf{v}_d(\vartheta_k^d)\in\mathbb{C}^{N_\mathrm{t}^d\times 1}\\
&=\frac{1}{\sqrt{N_\mathrm{t}^d}}\left[1\ \exp\{-\jmath\pi\vartheta_k^d\}\ \cdots\ \exp\{-\jmath\pi(N_\mathrm{t}^d-1)\vartheta_k^d\}\right]^T.
    \end{aligned}
    \end{equation}
Note that the propagation properties of the downlink channel in the space domain are reflected by the space angles $\vartheta_k^{\mathrm{x}}$ and $\vartheta_k^{\mathrm{y}}$ given by $\vartheta_k^{\mathrm{x}}=\sin{\theta_k^{\mathrm{y}}}\cos{\theta_k^{\mathrm{x}}}$ and $\vartheta_k^{\mathrm{y}}=\cos{\theta_k^{\mathrm{x}}}$, respectively \cite{you2020massive}, where $\theta_k^{\mathrm{x}}$ and $\theta_k^{\mathrm{y}}$ are the physical angles shown in Fig. \ref{COM_SYS}.
In addition, the channel gain $g_k(t,f)$ for UT $k$ is given by $g_k(t,f)\triangleq \sum_{l=1}^{L_{\rm k}}\alpha_{k,l}\exp\{\jmath 2\pi[t\nu_{k,l}^{\rm ut}-f\tau_{k,l}^{\rm ut}]\}$,
whose statistical properties are mainly determined by the location of the UTs \cite{you2020massive}.
Then, we model $g_k(t,f)$ to follow the Rician fading distribution with the Rician factor $\kappa_k$ and $\mathbb{E}\{|g_k(t,f)|^2\}=\gamma_k$.
In particular, the real and imaginary parts of $g_k(t,f)$ are
independently and identically distributed as $\mathcal{N}\left(\sqrt{\kappa_k\gamma_k/2(\kappa_k+1)}, \gamma_k/2(\kappa_k+1)\right)$.
Subsequently, with perfect time and frequency synchronization, the effective downlink channel vector between the transmitter and the $k$th UT is given by \cite{you2020massive,li2020downlink}
    \begin{align}\label{eq:chmd}
    \mathbf{h}_k(t,f)=\mathbf{v}_kg_k(t,f).
    \end{align}
In this work, we exploit the sCSI at the transmitter for downlink precoding in massive MIMO LEO SATCOM, including the UPA response vector $\bv_k$ and the statistical information of the channel gain $g_k$.
Note that the space angles in $\bv_k$ can be assumed to be fixed during the time interval of interest
and they should be updated with significantly change of the positions of the UTs and LEO
satellite \cite{clerckx2013mimo}.
In the following, we investigate each frequency sampling point for every coherence time interval and omit the instant $t$ and frequency $f$ for brevity.
Moreover, the channel spatial covariance matrix of the $k$th UT is given by
$\bC_k=\mathbb{E}\left\{\bh_k\bh_k^H\right\}=\gamma_k\bv_k\bv_k^H$,
which depends on the channel parameters $\{\gamma_k, \bv_k\}_{k=1}^K$.
\subsection{Transmission Model With Hybrid Precoding}

\begin{figure}[!t]
\centering
\subfloat[Fully connected architecture.]{\includegraphics[width=0.45\textwidth]{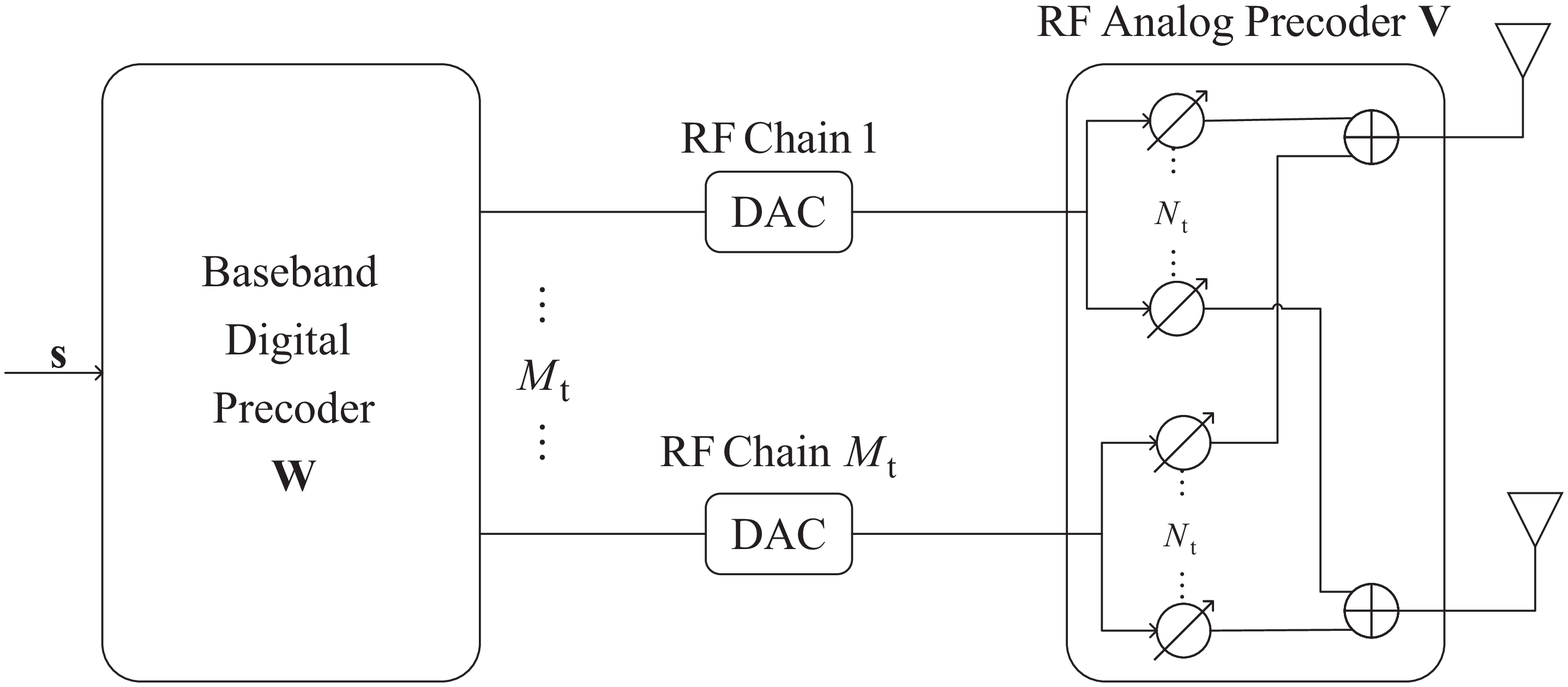}\label{framework1}}
\hfill
\subfloat[Partially connected architecture.]{\includegraphics[width=0.45\textwidth]{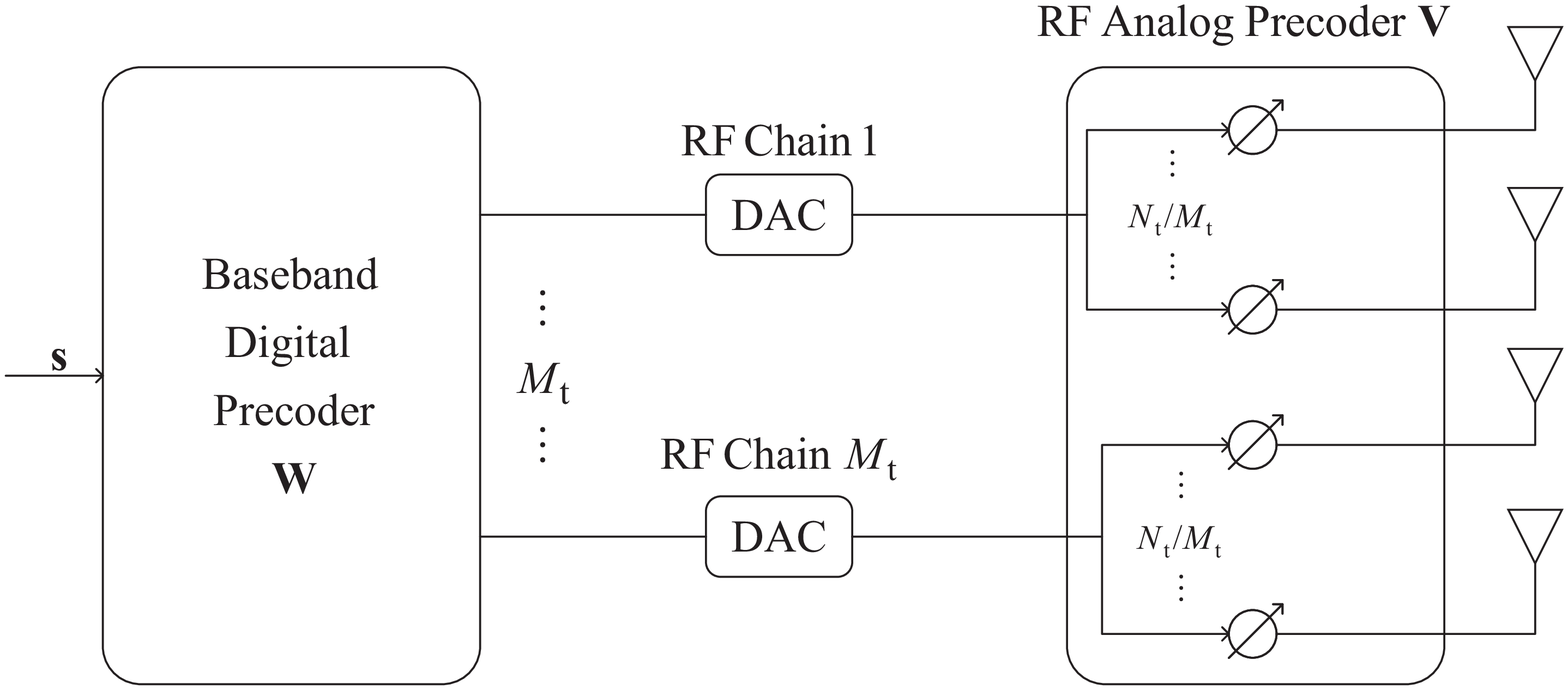}\label{framework2}}
\caption{Two architectures of hybrid precoding using different mapping strategies: each RF chain is connected to $N_{\mathrm{t}}$ antennas in (a) and $N_{\mathrm{t}}/M_{\mathrm{t}}$ antennas in (b).}
\label{framework}
\end{figure}

With proper time and frequency synchronization \cite{you2020massive}, the signal received by UT $k\in \{1,\ldots,K\}$ can be represented as
    \begin{equation}\label{eq:rcsg}
    \begin{aligned}
    y_k=\mathbf{h}_k^H\sum_{i=1}^K\mathbf{b}_is_i+n_k
    =\mathbf{h}_k^H\mathbf{b}_ks_k+\mathbf{h}_k^H\sum_{i\neq k}\mathbf{b}_is_i+n_k,
    \end{aligned}
    \end{equation}
where $n_k\sim \mathcal{CN}(0,N_0)$ is the circular symmetric complex-valued additive Gaussian white noise variable of the $k$th UT with a mean zero and variance $N_0$, $s_k$ is the signal for UT $k$ with mean 0 and variance 1, and $\mathbf{b}_k\in \mathbb{C}^{N_\mathrm{t}\times 1}$ is the precoding vector for UT $k$. Based on the applied hybrid architecture, as shown in Fig. \ref{framework}, the precoding vector can be factorized as $\mathbf{b}_k=\mathbf{Vw}_k$ \cite{yu2016alternating}. We denote the number of RF chains as $M_\mathrm{t}\left(K\leq M_\mathrm{t}\leq N_\mathrm{t}\right)$. Then, $\mathbf{V}$ represents the $N_\mathrm{t}\times M_\mathrm{t}$ RF analog precoder implemented through the phase shift network and $\mathbf{w}_k$ is the $M_\mathrm{t}\times 1$ baseband digital one.

In this paper, we assume that the phase shift network that maps the signal vector from the RF chains to the antennas is implemented via the two most common approaches in the literature, namely the fully and the partially connected architectures, as illustrated in Fig. \ref{framework}(a) and \ref{framework}(b), respectively. A similar model can be derived for the receiver's side via replacing the power amplifiers with low noise amplifiers (LNA) and the digital-to-analog converters (DACs) with analog-to-digital converters (ADCs).

\subsubsection{Baseband Digital Precoder}
The transmission vector after baseband digital precoding is defined as $\bu\triangleq\bW\bs$,
where $\mathbf{u}=\left[u_1,u_2,\ldots,u_{M_\mathrm{t}}\right]^T\in \mathbb{C}^{M_\mathrm{t}\times 1}$ and $\bs=[s_1,s_2,\ldots,s_K]^T\in \mathbb{C}^{K\times 1}$ is the transmit data signal with $\mathbb{E}\{\bs\bs^H\}=\bI$, processed by a linear baseband digital precoder $\mathbf{W}=[\mathbf{w}_1,\mathbf{w}_2,\ldots,\mathbf{w}_K]\in \mathbb{C}^{M_\mathrm{t}\times K}$.
\subsubsection{RF Analog Precoder}
The signal vector $\mathbf{u}$ can be precoded into the final transmission vector $\mathbf{V}\mathbf{u}$ through the analog precoder $\mathbf{V}\in \mathbb{C}^{N_\mathrm{t}\times M_\mathrm{t}}$ and its phase rotation is implemented by a phase shift network. We assume that the phase shifters are implemented with $\lambda$-bit resolution through a uniform quantizer and the step size of which is given by $\Delta=2\pi/2^{\lambda}$.

\begin{itemize}
  \item For the fully connected architecture, each entry of the analog precoder $\bV$ is unit-modulus, i.e., $|\bV_{i,j}|=1$.
  \item For the partially connected architecture, the structure of $\bV$ can be depicted as
    \begin{equation}\label{eq:pchv}
    \begin{aligned}
    \mathbf{V}=\blkdiag{\bp_1,\bp_2,\ldots,\bp_{M_{\rm t}}},
    \end{aligned}
    \end{equation}
where $\mathbf{p}_i\in\bbC^{N_\mathrm{t}/M_\mathrm{t}\times1}$ and each element of $\mathbf{p}_i$ is unit-modulus.
\end{itemize}

With respect to the feasible set of the above phase shifts, we take the following assumptions into account:
\begin{itemize}
  \item Continuous Phase Shifter (CPS): The resolutions of the phase shifters for implementing the analog precoder are all ideally infinite, i.e., $\lambda=\infty$. Then, the analog precoder $\bV$ can be modeled as
      \begin{align}
      \mathcal{S}_{\rm CPS, 1}&\triangleq\left\{\mathbf{V}\big | \bV_{i,j}=\exp\left\{\jmath\theta\right\},\ \theta\in [0,2\pi),\ \forall i,j\right\},\label{eq:cfcv}\\
      \mathcal{S}_{\rm CPS, 2}&\triangleq\left\{\mathbf{V}\big | \bV_{i,j}=\exp\left\{\jmath\theta\right\},\ \theta\in [0,2\pi),\ \right.\notag\\
      &\qquad \qquad \qquad \qquad \ \left. \forall i,\ \forall j=\left\lceil i\frac{M_{\rm t}}{N_{\rm t}}\right\rceil\right\},\label{eq:cpcv}
      \end{align}
for the fully and the partially connected architectures, respectively.
      \item Discrete Phase Shifter (DPS): In this case, we assume that the resolutions of all the phase shifters are equal and finite, and each phase shifter can only take values from $L = 2^{\lambda}$ different elements uniformly distributed in the interval $[0,2\pi)$. Then, the values of the entries of the analog precoder $\bV$ satisfy
      \begin{align}
            \mathcal{S}_{\rm DPS, 1}&\triangleq\left\{\mathbf{V}\Big | \bV_{i,j}=\exp\left\{\jmath\left(\frac{2\pi}{L}m+\frac{\pi}{L}\right)\right\},\ \right.\notag \\
            &\qquad \left. \forall i,j,\ m=0,1,\ldots,L-1\right\}, \label{eq:dfcv}\\
            \mathcal{S}_{\rm DPS, 2}&\triangleq\left\{\mathbf{V}\Big |\bV_{i,j}=\exp\left\{\jmath\left(\frac{2\pi}{L}m+\frac{\pi}{L}\right)\right\},\ \right.\notag \\
            &\qquad \left. \forall i, \ \forall j=\left\lceil i\frac{M_{\rm t}}{N_{\rm t}}\right\rceil,\ m=0,1,\ldots,L-1\right\},\label{eq:dpcv}
      \end{align}
      for the fully and the partially connected architectures, respectively.
\end{itemize}

\subsection{Power Consumption Model}

The total power consumption of the considered LEO SATCOM system can be modeled as \cite{arora19hybrid,cui2004energy,mendezrial2016hybrid}
    \begin{equation}\label{eq:totp}
    \begin{aligned}
    P^{\mathrm{total}}=\sum_{k=1}^K\xi ||\bb_k||_2^2+P_\mathrm{t},
    \end{aligned}
    \end{equation}
where $1/\xi$ and $P_\mathrm{t}$ denote the efficiency of transmit power amplifiers and the power consumed by the transmitter at the satellite, respectively. In the following, we present more details of the power consumption models at the transmitter.

For the downlink transmitter at the LEO satellite, the power consumption with a fully connected hybrid precoding architecture can be modeled as
    \begin{align}\label{eq:fptd}
    P_\mathrm{t}(M_\mathrm{t},\lambda)&=N_\mathrm{t}M_\mathrm{t}P_{\mathrm{Ps}}(\lambda)+M_\mathrm{t}P_{\mathrm{RFC}}+P_{\mathrm{LO}}+P_{\mathrm{BB}},
    \end{align}
while the expression can be converted into
    \begin{align}\label{eq:pptd}
    P_\mathrm{t}(M_\mathrm{t},\lambda)=N_\mathrm{t}P_{\mathrm{Ps}}(\lambda)+M_\mathrm{t}P_{\mathrm{RFC}}+P_{\mathrm{LO}}+P_{\mathrm{BB}},
    \end{align}
for a partially connected one, where $P_{\mathrm{Ps}}(\lambda)$ denotes the power consumption of the phase shifters depending on the resolution of the quantized phases \cite{mendezrial2016hybrid, chen2018low}, $P_{\mathrm{LO}}$ and $P_{\mathrm{BB}}$ are the power consumed by the local oscillator and baseband digital precoder, respectively. The power consumed by each RF chain can be further written as $P_{\mathrm{RFC}}=P_{\mathrm{DAC}}+P_{\mathrm{mixer}}+P_{\mathrm{LPF}}+P_{\mathrm{BBA}}$,
where $P_{\mathrm{DAC}}$, $P_{\mathrm{mixer}}$, $P_{\mathrm{LPF}}$, and $P_{\mathrm{BBA}}$ denote the power consumption of a single DAC, mixer, low pass filter, and baseband amplifier, respectively.
In addition, for the case of a fully digital transmitter at the LEO satellite, the total power consumption can be written as
    \begin{align}\label{eq:fdpt}
    P_\mathrm{t}=N_\mathrm{t}P_{\mathrm{RFC}}+P_{\mathrm{LO}}+P_{\mathrm{BB}}.
    \end{align}
\subsection{Problem Formulation}


According to the received signal model in \eqref{eq:rcsg}, the SINR of UT $k$ in the downlink transmission can be defined as
    \begin{align}\label{eq:dlsinr}
    \sinr_{k}\triangleq \frac{|\mathbf{b}_k^H\mathbf{h}_k|^2}{\sum_{\ell\neq k}|\mathbf{b}_{\ell}^H\mathbf{h}_k|^2+N_0}.
    \end{align}
Note that the instantaneous SINR expression in \eqref{eq:dlsinr} is achieved if $|\mathbf{b}_k^H\mathbf{h}_k|$ is known at the receiver of UT $k$. Let us now assume long-term sCSI knowledge at the transmitter, including $\bv_k$ and the statistics of $g_k,\forall k$, for the downlink precoding regime in the SATCOM system under consideration.
Then, the average data rate of the $k$th UT at the transmitter is given by $R_k=\mathbb{E}\{\log(1 + \sinr_k)\}$,
and the total achievable system EE can be defined as \cite{tsinos2017energy}
    \begin{align}\label{eq:eedf}
    {\rm EE}=\frac{B_{\rm w}\sum_{k=1}^KR_k}{P^{\mathrm{total}}},
    \end{align}
where $B_{\rm w}$ is the system bandwidth.



The objective of our work is to maximize the total achievable system EE for hybrid precoding in the massive MIMO LEO SATCOM system, which can be formulated as
    \begin{subequations}\label{eq:eemn}
    \begin{align}
    \mathcal{P}_1:\mathop{\mathrm{maximize}}\limits_{\bV,\{\mathbf{w}_k\}_{k=1}^K}&\ \ \frac{B_{\rm w}\sum_{k=1}^K R_k}{P^{\mathrm{total}}} \label{eq:eemna}\\
    \mathrm{s.t.}&\ \ \sum_{k=1}^K||\bV\bw_k||_2^2\leq P,\label{eq:eemnb}\\
    \ \ \ \ &\ \ \mathbf{V}\in \mathcal{S},\label{eq:eemnc}
    \end{align}
    \end{subequations}
where $\mathcal{S}\in \{\mathcal{S}_{\rm CPS,1}, \mathcal{S}_{\rm CPS,2}, \mathcal{S}_{\rm DPS,1}, \mathcal{S}_{\rm DPS,2}\}$ is defined in \eqref{eq:cfcv}--\eqref{eq:dpcv} for the different considered analog precoding architectures.

Note that problem $\mathcal{P}_1$ in \eqref{eq:eemn} is a fractional nonconvex problem and $P$ is a total downlink transmit power constraint. Due to the coupled digital and analog precoders in \eqref{eq:eemnb} and the nonconvex constraint in \eqref{eq:eemnc}, this problem is in general difficult to tackle. To that end, we develop a tractable solution by setting $\bb_k = \bV_k\bw_k$. We first focus on the fully digital equivalent problem. Then, the hybrid precoders are derived as the minimizers of the Euclidean distance to the obtained fully digital solution \cite{el2014spatially}.

\section{Optimization of Fully Digital Problem}\label{sec:fullydigital_solution}

In this section, we develop a tractable solution to the fully digital equivalent problem. First, we adopt a closed-form upper bound of the average rate. Subsequently, Dinkelbach's algorithm and the WMMSE method are adopted to convert the nonconvex problem into several convex subproblems which can be iteratively handled.

\subsection{Upper Bound of the Average Rate}

The average rate function $R_k$ does not have an explicit expression in general, which is usually tricky to handle. A Monte Carlo method can be adopted to estimate the required value but its computational complexity is high.
Then, we substitute the ergodic data rate $R_k$ with an upper bound in the following optimization procedure, which is given by
    \begin{equation}\label{eq:exrt}
    \begin{aligned}
    R_k\leq \bar{R}_k
    &\triangleq\log\left(1+\frac{|\mathbf{b}_k^H\mathbf{v}_k|^2\mathbb{E}\left\{|g_k|^2\right\}}{\sum_{{\ell}\neq k}|\mathbf{b}_{\ell}^H\mathbf{v}_k|^2\mathbb{E}\left\{|g_k|^2\right\}+N_0}\right)\\
    &=\log\left(1+\frac{\gamma_k|\mathbf{v}_k^H\mathbf{b}_k|^2}{\sum_{{\ell}\neq k}\gamma_k|\mathbf{v}_k^H\mathbf{b}_{\ell}|^2+N_0}\right).
    \end{aligned}
    \end{equation}
\emph{Proof:} Please refer to \appref{app:a}.

The upper bound in \eqref{eq:exrt} is tight in typical scenarios, which will be verified in the simulations in \ref{sec:sim}.
Furthermore, with the utilization of the upper bound for the ergodic rate, the precoders are designed based on the channel parameters $\left\{\bv_k\right\}_{k=1}^K$ and $\left\{\gamma_k\right\}_{k=1}^K$.
Note that the array response vectors $\left\{\bv_k\right\}_{k=1}^K$  are determined by the space angle pairs $\left\{\left(\vartheta_k^{\rm x},\vartheta_k^{\rm y}\right)\right\}_{k=1}^K$, which can be computed from the position knowledge of the satellite and the UTs with the assistance of e.g., global positioning system (GPS) \cite{spilker1978gps,kaplan2006global}.
Besides, the average channel power parameters $\left\{\gamma_k\right\}_{k=1}^K$ are generally slow-varying and it is not very difficult to estimate them. For example, they can be estimated by exploiting the reciprocity between the uplink and downlink channels of the considered system \cite{sun2015beam}.
Then, it is worth noting that the sCSI parameters, i.e., $\left\{\gamma_k,\vartheta_k^{\rm x},\vartheta_k^{\rm y}\right\}_{k=1}^K$, are generally slow-varying and low dimensional, thus are not difficult to estimate, which can be incorporated in the precoding design to reduce the complexity.

\subsection{Dinkelbach's Algorithm and WMMSE Method}\label{sec:fullydigital_solution_alg}


With Dinkelbach's algorithm \cite{zapponealessio2015energy}, we handle the fully digital conversion of $\mathcal{P}_1$ iteratively through a series of subproblems, which converges to the solution of the original problem \cite{rodenas1999extensions}. Then, we represent $n$ as the index and denote $\rho_n$ as the auxiliary variable in the $n$th iteration. Subsequently, the $n$th subproblem is shown as
%
    \begin{subequations}
    \begin{align}
    \mathcal{P}_2^n: \mathop{\mathrm{maximize}}\limits_{\{\mathbf{b}_k^n\}_{k=1}^K, \rho_n}&\ \ F(\rho_n)=B_{\rm w}\sum_{k=1}^K\bar{R}_k-\rho_n P^{\mathrm{total}}\label{eq:dbao}\\
    \mathrm{s.t.}&\ \ \sum_{k=1}^K||\bb_k^n||_2^2\leq P.\label{eq:dbac}
    \end{align}
    \end{subequations}
For subproblem $\mathcal{P}_2^n$, the variables $\{\mathbf{b}_k^n\}_{k=1}^K$ and $\rho_n$ are optimized in an iterative way. Specifically, with given $\{\mathbf{b}_k^n\}_{k=1}^K$, the value of $\rho_{n+1}$ for problem $P_2^{n+1}$ is given by \cite{zapponealessio2015energy}
    \begin{align}
\rho_{n+1}=\frac{B_{\rm w}\sum_{k=1}^K\bar{R}_k}{P^{\mathrm{total}}}.
    \end{align}

In the following, we address problem $\mathcal{P}_2^n$ with a given $\rho_n$. Since the procedure is the same for each subproblem, we omit the index $n$ for convenience.
Specifically, utilizing the WMMSE method \cite{shi2011an}, we first derive an equivalent minimization problem for $\mathcal{P}_2$, which is given by
    \begin{equation}\label{eq:wmse}
    \begin{aligned}
    \mathcal{P}_3: \mathop{\mathrm{minimize}}\limits_{\{\mathbf{b}_k, \omega_k, u_k\}_{k=1}^K}&B_{\rm w}\sum_{k=1}^K(\omega_ke_k-\log\omega_k)+\rho P^{\mathrm{total}}\\
    {\rm s.t.}\ \ &\sum_{k=1}^K||\bb_k||_2^2\leq P,
    \end{aligned}
    \end{equation}
where $e_k$ is the mean-square estimation error of the transmit signal $s_k$ and the estimated signal considering the linear receiver $u_k$ \cite{shi2011an}, given by
    \begin{align}\label{eq:msee}
    e_k\triangleq |u_k\sqrt{\gamma_k}\bv_k^H\bb_k-1|^2+\sum_{i\neq k}\gamma_k|u_k\bv_k^H\bb_i|^2+N_0|u_k|^2.
    \end{align}
Note that the three variables of problem $\mathcal{P}_3$ in (24) are tightly decoupled, which is in general challenging to tackle. To that end, the block coordinate descent method is adopted to handle problem $\mathcal{P}_3$ by assuming that two of the three variables are fixed and update the remaining one \cite{shi2011an,bertsekas1997nonlinear}, whose convergence is guaranteed when the optimization problem for each variable is solved \cite{bertsekas1997nonlinear}.
Specifically, by assuming that two of the three variables are fixed, the remaining one can be updated. The expressions for updating $\omega_k$ and $u_k$ are respectively given by
    \begin{subequations}\label{eq:opwu}
    \begin{align}
    u_k&=\frac{\sqrt{\gamma_k}\bv_k^H\bb_k}{\sum_{i=1}^K\gamma_k|\bv_k^H\bb_i|^2+N_0},\label{eq:opu}\\
    \omega_k&=e_k^{-1}.\label{eq:opw}
    \end{align}
    \end{subequations}
With $\omega_k$ and $u_k$ being fixed, problem $\mathcal{P}_3$ can be simplified after dropping terms independent to $\bb_k$ as follows
    \begin{subequations}\label{eq:cqop}
    \begin{align}
    \mathcal{P}_4: \mathop{\mathrm{minimize}}\limits_{\{\mathbf{b}_k\}_{k=1}^K}&\ \ B_{\rm w}\sum_{k=1}^K\omega_k\left(|u_k\sqrt{\gamma_k}\bv_k^H\bb_k-1|^2\right.\notag\\
    &\qquad \left. +\sum_{i\neq k}\gamma_k|u_k\bv_k^H\bb_i|^2\right)+\rho \sum_{k=1}^K\xi ||\bb_k||_2^2\label{eq:cqopa}\\
    {\rm s.t.}&\ \ \sum_{k=1}^K||\bb_k||_2^2\leq P,\label{eq:cqopb}
    \end{align}
    \end{subequations}
which is a convex quadratic optimization problem and can be directly solved by using standard convex optimization algorithms. However, it has high computational complexity, so we use the Lagrange multipliers method \cite{boyd2004convex} which will be illustrated in the following. In particular, a corresponding Lagrange function can be obtained with a Lagrange multiplier $a$ attributed to the power budget constraint \eqref{eq:cqopb}, which is given by
    \begin{align}\label{eq:lfun}
    &L(\{\bb_k\}_{k=1}^K,a)\notag\\
    =&B_{\rm w}\sum_{k=1}^K\omega_k\left(|u_k\sqrt{\gamma_k}\bv_k^H\bb_k-1|^2+\sum_{i\neq k}\gamma_k|u_k\bv_k^H\bb_i|^2\right)\notag\\
    &\qquad +\rho \sum_{k=1}^K\xi ||\bb_k||_2^2+a\left(\sum_{k=1}^K||\bb_k||_2^2-P\right).
    \end{align}
Then, the KKT conditions for problem $\mathcal{P}_4$ in \eqref{eq:cqop} are given by \cite{boyd2004convex}
\begin{subequations}
\begin{align}
&\sum_{k=1}^K||\bb_k^{\rm opt}||_2^2\leq P, \\
&a^{\rm opt}\geq 0,\\
&a^{\rm opt}\left(\sum_{k=1}^K||\bb_k^{\rm opt}||_2^2-P\right)=0,\label{eq:comple_sla_cond}\\
&\left(\bb_k^{\rm opt}\right)^H\sum_{\ell =1}^KB_{\rm w}\omega_{\ell}\gamma_{\ell}\abs{u_{\ell}}^2\bv_{\ell}\bv_{\ell}^H-\omega_ku_k\sqrt{\gamma_k}\bv_k^H \xi\notag\\
&\qquad \qquad \qquad \qquad \qquad +(\rho+a^{\rm opt})\left(\bb_k^{\rm opt}\right)^H=0 \label{eq:optbepr}.
\end{align}
\end{subequations}
Based on Eq. \eqref{eq:optbepr}, we can get
    \begin{equation}\label{eq:lfun}
    \begin{aligned}
    \bb_k^{\rm opt}&=\omega_k(\sqrt{\gamma_k})u_k^{\ast}\left(\sum_{k=1}^KB_{\rm w}\omega_k|u_k|^2\gamma_k\bv_k\bv_k^H\right.\\
    &\qquad \qquad \qquad \qquad \qquad \left.+\left(\rho\xi+a^{\rm opt}\right)\bI\right)^{-1}\bv_k.
    \end{aligned}
    \end{equation}
The value of parameter $a^{\rm opt}$ can be determined through the following steps which ensure the satisfaction of the complementarity slackness condition \eqref{eq:comple_sla_cond}. In particular, if the term $\sum_{k=1}^KB_{\rm w}\omega_k|u_k|^2\gamma_k\bv_k\bv_k^H+\rho\xi\bI$ is invertible and $\sum_{k=1}^K\left\|\bb_k^{\rm opt}\right\|_2^2\leq P$, then $a^{\rm opt}=0$,
otherwise we must have
    \begin{equation}\label{eq:lcst}
    \begin{aligned}
    \sum_{k=1}^K||\bb_k^{\rm opt}||^2_2= P.
    \end{aligned}
    \end{equation}
Hence, $a^{\rm opt}$ can be derived via \eqref{eq:lcst} using the bisection method \cite{shi2011an}. The complete procedure is summarized in \textbf{Algorithm \ref{alg:algwmmse}}.

\begin{algorithm}
\caption{Precoder Design Algorithm for Problem \eqref{eq:eemn}}
\label{alg:algwmmse}
\begin{algorithmic}[1]
\REQUIRE Thresholds $\epsilon_1, \epsilon_2>0$, $n=0$, $\rho_n=0$.
\ENSURE Precoding vector $\bb_k$.
\WHILE{$F(\rho_n)>\epsilon_1$}
\STATE Initialize $\bb_k$ such that $||\bb_k||^2_2=P/K$.
\REPEAT
\STATE Update
\begin{align}
\omega_k' &= \omega_k,\notag\\
u_k &= \sqrt{\gamma_k}\bv_k^H\bb_k\left(\sum_{i=1}^K\gamma_k|\bv_k^H\bb_i|^2+N_0\right)^{-1},\notag\\
\omega_k&=\left(|u_k\sqrt{\gamma_k}\bv_k^H\bb_k-1|^2+\sum_{i\neq k}\gamma_k|u_k\bv_k^H\bb_i|^2\right.\notag\\
&\qquad \qquad \qquad \qquad \qquad \qquad \left.+N_0|u_k|^2\right)^{-1},\notag\\
\bb_k^{\rm opt}&=\omega_k(\sqrt{\gamma_k})u_k^{\ast}\left(\sum_{k=1}^KB_{\rm w}\omega_k|u_k|^2\gamma_k\bv_k\bv_k^H\right.\notag\\
&\qquad \qquad \qquad \qquad \left.+\left(\rho\xi+a^{\rm opt}\right)\bI\right)^{-1}\bv_k.\notag
\end{align}
\UNTIL{$|\sum_k\log\omega_k-\sum_k\log\omega_k'|<\epsilon_2$}.
\STATE $F(\rho_n)=(B_{\rm w}\sum_{k=1}^K\bar{R}_k-\rho_n P^{\mathrm{total}})|_{\bb_k=\bb_k^{\rm opt}}$.
\STATE $\rho_{n+1}=\frac{B_{\rm w}\sum_{k=1}^K\bar{R}_k}{P^{\mathrm{total}}}\Big |_{\bb_k=\bb_k^{\rm opt}}$.
\STATE $n=n+1$.
\ENDWHILE
\end{algorithmic}
\end{algorithm}

\subsection{Hybrid Precoding Optimization}\label{sec:fullydigital_solution_hyp}

In this subsection, we aim at the design of a hybrid precoder $\bV\bw_k$ that approximates the digital precoding vector $\bb_k$  derived in Subsection \ref{sec:fullydigital_solution_alg}. In particular, the hybrid precoder $\mathbf{V}$ and $\mathbf{w}_k$ can be found by minimizing the Euclidean distance between $\mathbf{Vw}_k$ and $\mathbf{b}_k$ as follows \cite{arora2019hybrid}
    \begin{subequations}\label{eq:mind}
    \begin{align}
    \mathcal{Q}_1: \mathop{\mathrm{minimize}}\limits_{\mathbf{W},\mathbf{V}}&\ \ ||\mathbf{B}-\mathbf{VW}||_F^2\label{eq:minda}\\
    {\rm s.t.}&\ \ \mathbf{V}\in \mathcal{S},\label{eq:mindb}\\
    & \ \ ||\mathbf{B}||_F^2=||\mathbf{VW}||_F^2,\label{eq:mindc}
    \end{align}
    \end{subequations}
where $\mathbf{B}=[\mathbf{b}_1,\mathbf{b}_2,\ldots,\mathbf{b}_K]$ and $\mathbf{W}=[\mathbf{w}_1,\mathbf{w}_2,\ldots,\mathbf{w}_K] $.

For problem $\mathcal{Q}_1$, an alternating algorithm can be applied to decouple the analog and digital precoders.
In that context, one precoder is assumed to be fixed while the other is optimized and vice-versa.
This procedure is repeated until the relative change in the value of the objective function is less than a certain threshold.
Note that most existing works focus on the infinite-resolution phase shift network. However, the analog precoder with continuous phase per element is not easy to implement in
practise and the corresponding method can not be directly applied to our work.
In our work, a finite-resolution phase shift network is considered and thus, the phase of each non-zero element of the analog precoder is discrete.
Then, we develop specific algorithms for the fully and the partially connected hybrid precoding architectures, which can be applied for both the finite- and infinite-resolution scenarios, in Sections \ref{sec:fullyconnected} and \ref{sec:partiallyconnected}, respectively.
\subsection{Convergence and Complexity Analysis}
It is shown above that \textbf{Algorithm \ref{alg:algwmmse}} is a combination of Dinkelbach's algorithm with the WMMSE method. The corresponding convergence analysis of Dinkelbach's algorithm can be found in \cite{zapponealessio2015energy}. Introducing an auxiliary index parameter, the original fractional programming problem can be converted into a series of subproblems. The overall computational complexity is determined by the following two factors. First, it is demonstrated in \cite{zapponealessio2015energy} that the sequence of subproblems in Dinkelbach's algorithm converges and the complexity can be estimated as $\mathcal{O}(I_{\rm D})$ with a small number of iterations, $I_{\rm D}$. Second, it comes to the complexity of each individual subproblem, tackling through the WMMSE method. The convergence of the WMMSE method is proved in \cite{shi2011an}.

Both the complexity for the updates of $u_k$ and $\omega_k$ through \eqref{eq:opu} and \eqref{eq:opw} in each iteration can be shown to be equal to $\mathcal{O}(KN^2_{\rm t})$. The computation of $\bb_k$ includes a bisection search process with a small number of iterations which can usually be ignored there. Thus, the complexity of \eqref{eq:lfun} is simplified to $\mathcal{O}(KN_{\rm t}^3)$. Assuming that the WMMSE method involves $I_{\rm M}$ iterations, the whole complexity of \textbf{Algorithm \ref{alg:algwmmse}} is approximately equal to $\mathcal{O}(I_{\rm D}I_{\rm M}(2KN^2_{\rm t}+KN_{\rm t}^3))$.

\section{Hybrid Precoding for Fully Connected Architecture}\label{sec:fullyconnected}

The hybrid precoding optimization problem has been put forward in Section \ref{sec:fullydigital_solution_hyp}. In this section, we investigate this problem for the fully connected architecture, and the case for the partially connected architecture will be investigated in the next section. Specifically, we first set a feasible analog precoder and derive a closed-form digital precoder. Then, for the given digital precoder, the analog precoder $\bV$ is optimized via the application of an inexact MM-based method.

\subsection{Digital Precoder Design}
By assuming a fixed analog precoder $\mathbf{V}$, the digital precoder design can be formulated as the following problem
    \begin{subequations}\label{eq:temp}
    \begin{align}
    \mathop{\mathrm{minimize}}\limits_{\mathbf{W}}&\ \ ||\mathbf{B}-\mathbf{VW}||_F^2\label{eq:tempa}\\
    {\rm s.t.}&\ \ ||\mathbf{B}||_F^2=||\mathbf{VW}||_F^2,\label{eq:tempb}
    \end{align}
    \end{subequations}
which is a nonconvex quadratically constrained quadratic program with the equality constraint in \eqref{eq:tempb}.
First, we drop this constraint and consider the corresponding unconstrained problem whose solution admits a closed-form that can be given by $\mathbf{W}=(\mathbf{V}^H\mathbf{V})^{-1}\mathbf{V}^H\mathbf{B}$.
Then, the expression is normalized to satisfy the equality constraint as
    \begin{equation}
    \begin{aligned}
    \mathbf{W}=\frac{||\mathbf{B}||_F}{||\mathbf{VW}||_F}\mathbf{W}.
    \end{aligned}
    \end{equation}
It is proven that the solution can make the value of the objective function sufficiently small \cite{yu2016alternating}.
\subsection{Analog Precoder Design}
With a fixed digital precoder $\mathbf{W}$, problem \eqref{eq:mind} can be simplified as follows
    \begin{subequations}\label{eq:edmv}
    \begin{align}
    \mathcal{Q}_2: \mathop{\mathrm{minimize}}\limits_{\mathbf{V}}&\ \ f\left(\mathbf{V}\right)=||\mathbf{B}-\mathbf{VW}||_F^2\\
    {\rm s.t.}&\ \ \mathbf{V}\in \mathcal{S}_1,
    \end{align}
    \end{subequations}
where $\mathcal{S}_1\in \{\mathcal{S}_{\rm CPS,1}, \mathcal{S}_{\rm DPS,1}\}$. The challenge of problem \eqref{eq:edmv} lies in the phase shift constraint for the analog precoder.
To make it more tractable, we introduce the following lemma,  which can be derived based on the results in \cite{shao2019framework}.

\begin{lemma}\label{lemma:eePenalty}
For some certain $\bar{\eta}>0$, an auxiliary $\eta>\bar{\eta}$ is introduced and problem \eqref{eq:edmv} can be equivalently converted to the following one:
\begin{equation}\label{eq:immm}
\begin{aligned}
\mathcal{Q}_3: \mathop{\mathrm{minimize}}\limits_{\mathbf{V}\in \tilde{\mathcal{S}}_1}\ \ F_{\eta}(\mathbf{V})\triangleq f(\mathbf{V})-\eta||\mathbf{V}||_F^2
\end{aligned}
\end{equation}
where $\tilde{\mathcal{S}}_1$ denotes the convex hull of $\mathcal{S}_1$.
\end{lemma}

Since $\mathcal{Q}_3$ is generally a difference of convex problem, which is nonconvex, we apply the MM method \cite{shao2019framework} to handle it. The basic idea of the MM method is to find a surrogate function $G_{\eta}(\mathbf{V}|\bar{\mathbf{V}})$ which maximizes $F_{\eta}(\mathbf{V})$ at $\bar{\mathbf{V}}$, i.e., it satisfies $F_{\eta}(\mathbf{V})\leq G_{\eta}(\mathbf{V}|\bar{\mathbf{V}}),\ \forall \mathbf{V},\bar{\mathbf{V}}$ and $F_{\eta}(\bar{\mathbf{V}})\leq G_{\eta}(\bar{\mathbf{V}}|\bar{\mathbf{V}}),\ \forall \bar{\mathbf{V}}$. Note that for any $\mathbf{V}$ and $\bar{\mathbf{V}}$, $||\mathbf{V}||_F^2\geq ||\bar{\mathbf{V}}||_F^2+2\langle\bar{\mathbf{V}},\mathbf{V}-\bar{\mathbf{V}}\rangle$, where $\langle\bar{\mathbf{V}},\mathbf{V}-\bar{\mathbf{V}}\rangle=
\Re\left({\rm Tr}\left\{\bar{\mathbf{V}}^H\left(\mathbf{V}-\bar{\mathbf{V}}\right)\right\}\right)$. Hence, we have
    \begin{equation}
    \begin{aligned}
    F_{\eta}(\mathbf{V})\leq  f(\mathbf{V})-\eta||\bar{\mathbf{V}}||_F^2-2\eta\langle\bar{\mathbf{V}},\mathbf{V}-\bar{\mathbf{V}}\rangle\triangleq G_{\eta}(\mathbf{V}|\bar{\mathbf{V}}).
    \end{aligned}
    \end{equation}
It is easy to check that $G_{\eta}(\mathbf{V}|\bar{\mathbf{V}})$ is the majorant of $F_{\eta}(\mathbf{V})$ at $\bar{\mathbf{V}}$ for problem $\mathcal{Q}_3$.  Thus, we tackle problem $\mathcal{Q}_3$ by repeatedly applying \cite{shao2019framework}
    \begin{equation}\label{eq:mmre}
    \begin{aligned}
    \mathbf{V}^{t+1}={\rm arg}\min_{\mathbf{V}\in \tilde{\mathcal{S}}_1}G_{\eta}(\mathbf{V}|\mathbf{V}^{t}),\ t=0,1,2,\ldots,
    \end{aligned}
    \end{equation}
until convergence.

In the following, we investigate how to perform the MM update in \eqref{eq:mmre} in a more efficient way. Note that problem \eqref{eq:mmre} is smooth with the convex hull of $\mathcal{S}_1$ and
    \begin{equation}
    \begin{aligned}
\nabla_{\mathbf{V}}G_{\eta}(\bar{\mathbf{V}}|\bar{\mathbf{V}})=&\nabla_{\mathbf{V}}F_{\eta}(\bar{\mathbf{V}})\\
=&2\left(\bar{\mathbf{V}}\mathbf{WW}^H-\mathbf{B}\mathbf{W}^H-\eta\bar{\mathbf{V}}\right),
    \end{aligned}
    \end{equation}
and we can handle it by employing the APG method \cite{shao2019framework}. Let $n=0,\hat{\mathbf{V}}^{n-1}=\hat{\mathbf{V}}^n=\mathbf{V}^{t}$ and  repeatedly perform the following iterations
    \begin{equation}\label{eq:inmm}
    \begin{aligned}
    \hat{\mathbf{V}}^{n+1}=\Pi_{\tilde{\mathcal{S}}_1}\left(\mathbf{Z}^n-\frac{1}{\beta_n}\nabla_{\mathbf{V}}G_{\eta}(\mathbf{Z}^n|\mathbf{V}^t)\right),
    n=0,1,2,\ldots,
    \end{aligned}
    \end{equation}
with the final $\hat{\mathbf{V}}^{n+1}$ as $\mathbf{V}^{t+1}$. In \eqref{eq:mmre}, $1/\beta_n>0$ is the chosen step size which can be found through a backtracking line search method \cite{shao2019framework} to satisfy the following so-called descent property which is expressed as
    \begin{equation}
    \begin{aligned}
    &G_{\eta}(\hat{\mathbf{V}}^{n+1}|\mathbf{V}^{t})\\
    \leq& G_{\eta}(\mathbf{Z}^n|\mathbf{V}^{t})+\langle\nabla_{\mathbf{V}}G_{\eta}(\mathbf{Z}^n|\mathbf{V}^{t}),\hat{\mathbf{V}}^{n+1}-\mathbf{Z}^n\rangle\\
    &\qquad \qquad \qquad \qquad \qquad \qquad+\frac{\beta_n}{2}||\hat{\mathbf{V}}^{n+1}-\mathbf{Z}^n||_F^2.
    \end{aligned}
    \end{equation}

However, it requires several iterations to find an exact solution of problem \eqref{eq:mmre} through the APG method, resulting in high computational complexity. Thus, we adopt an inexact MM update with only one APG step performed at each iteration \cite{cui2019hybrid}. Specifically,
    \begin{equation}
    \begin{aligned}
    \mathbf{V}^{n+1}=\Pi_{\tilde{\mathcal{S}}_1}\left(\mathbf{Z}^n-\frac{1}{\beta_n}\nabla_{\mathbf{V}}G_{\eta}(\mathbf{Z}^n|\mathbf{V}^n)\right),
    n=0,1,2,\ldots,
    \end{aligned}
    \end{equation}
where $\mathbf{Z}^n$ is an extrapolated point given by $\mathbf{Z}^n=\mathbf{V}^n+\zeta_n(\mathbf{V}^n-\mathbf{V}^{n-1})$,
with
    \begin{equation}
    \begin{aligned}
    \zeta_n=\frac{\alpha_{n-1}-1}{\alpha_n},\alpha_n=\frac{1+\sqrt{1+4\alpha_{n-1}^2}}{2},\alpha_{-1}=0,
    \end{aligned}
    \end{equation}
and $\Pi_{\tilde{\mathcal{S}}_1}(\bar{\mathbf{V}})$ denotes the element-wise projection of $\bar{\mathbf{V}}$ onto set $\tilde{\mathcal{S}}_1$. In particular, for the DPS case and some $v\in \mathbb{C}$, it can be calculated as follows
    \begin{equation}
    \begin{aligned}
    &\Pi_{\tilde{\mathcal{S}}_{\rm DPS,1}}(v)\\
=&\exp\left(\jmath\frac{2\pi m}{L}\right)\left(\left[\Re(\tilde{v})\right]_{0}^{\cos(\pi/L)}+\jmath\left[\Im(\tilde{v})\right]_{-\sin(\pi/L)}^{\sin(\pi/L)}\right),
    \end{aligned}
    \end{equation}
where $m=\lfloor\frac{\angle v+\pi/L}{2\pi/L}\rfloor,\tilde{v}=v \exp\left\{-\jmath\frac{2\pi m}{L}\right\},[x]_a^b=\min\{b,\max\{x,a\}\}$. In addition, for the CPS case,
    \begin{equation}
    \begin{aligned}
    \Pi_{\tilde{\mathcal{S}}_{\rm CPS,1}}(v)=\left\{
                                            \begin{array}{ll}
                                              v, & \text{if}\ v\leq 1 \\
                                              v/|v|, & \text{otherwise}.
                                            \end{array}
                                          \right.
    \end{aligned}
    \end{equation}
The complete procedure for problem $\mathcal{P}_1$ is shown in \textbf{Algorithm \ref{alg:algaadp}} and Step 3 of which is performed through the inexact MM method, which is summarized in \textbf{Algorithm \ref{alg:algiemm}}.


\begin{algorithm}[!t]
\caption{\textbf{A}lternating \textbf{I}nexact \textbf{MM} Based Hybrid \textbf{A}nalog/\textbf{D}igital \textbf{P}recoding (AIM-ADP)}
\label{alg:algaadp}
\begin{algorithmic}[1]
\REQUIRE A feasible point $\mathbf{V}\in \tilde{\mathcal{S}}_1$.
\REPEAT
\STATE
Fixed $\mathbf{V}$, find $\mathbf{W}=\mathbf{V}^H(\mathbf{V}\mathbf{V}^H)^{-1}\mathbf{B}$.
\STATE
Fixed $\mathbf{W}$, find $\mathbf{V}$ through \textbf{\alref{alg:algiemm}}.
\UNTIL The relative change of the objective value is less than a certain threshold.
\STATE The digital precoder is normalized as $\displaystyle \mathbf{W}=\frac{||\mathbf{B}||_F}{||\mathbf{VW}||_F}\mathbf{W}$.
\end{algorithmic}
\end{algorithm}

\begin{algorithm}[!t]
\caption{Inexact MM Method for Problem \eqref{eq:mmre}}
\label{alg:algiemm}
\begin{algorithmic}[1]
\REQUIRE Constant integers $J\geq 1$ and $c>1$, a feasible point $\mathbf{V}^0\in \tilde{\mathcal{S}}_1$, an initial penalty $\eta>0$ and its upper bound $\eta_{\mathrm{upp}}>0$ and threshold $\epsilon>0$.
\STATE $\mathbf{V}^{-1}=\mathbf{V}^{0}$.
\STATE $n=0$.
\REPEAT
\STATE Update
    \begin{align}
\mathbf{Z}^n&=\mathbf{V}^n+\zeta_n(\mathbf{V}^n-\mathbf{V}^{n-1})\notag,\\
\mathbf{V}^{n+1}&=\Pi_{\tilde{\mathcal{S}}}\left(\mathbf{Z}^n-\frac{1}{\beta_n}\nabla_{\mathbf{V}}G_{\eta}(\mathbf{Z}^n|\mathbf{V}^n)\right).\notag
    \end{align}
\STATE
Update $\eta=\eta c$ every $J$ iterations or when $||\mathbf{V}^{n+1}-\mathbf{V}^{n}||_F<\epsilon$.
\STATE
$n=n+1$.
\UNTIL $\eta>\eta_{\mathrm{upp}}$.
\end{algorithmic}
\end{algorithm}

\subsection{Convergence and Complexity Analysis}
\textbf{Algorithm \ref{alg:algaadp}} combines an alternating minimization framework and an inexact MM method based on the APG update steps. The algorithm is guaranteed to converge to a feasible point due to the iteratively minimization of the objective function at each iteration in Steps 2 and 3 and the non-negative characteristics of the objective function. In addition, the convergence analysis for the inexact MM method can be referred to \cite{shao2019framework}.

In each iteration, the complexity to calculate a pseudo-inverse matrix in the computation of the digital precoder in Step 2 is $\mathcal{O}(N_{\rm t}^3)$. For the analog precoder design, the complexity of each gradient update is given by $\mathcal{O}(KN_{\rm t}M_{\rm t})$. Based on the above analysis and assuming that there are $I_{\rm A}$ and $J$ iterations in the alternating minimization framework and \textbf{Algorithm 3}, respectively, the overall complexity of \textbf{Algorithm \ref{alg:algaadp}} can be estimated as $\mathcal{O}(I_{\rm A}(JKN_{\rm t}M_{\rm t}+N_{\rm t}^3))$.

\section{Hybrid Precoding for Partially Connected Architecture}\label{sec:partiallyconnected}

In this section, we investigate the partially connected architecture for hybrid precoding in LEO SATCOM.
Due to the block diagonal structure of the analog precoding matrix $\bV$ in this case, the digital baseband precoder can be derived through variable projection. Then, the analog precoder can be obtained through the inexact MM method similarly as that developed in the above section after some modifications.

\subsection{Problem Formulation}

Observing that each row of $\bW$ is multiplied by a corresponding nonzero entry from $\bV$ and therefore, the power constraint \eqref{eq:mindc} in $\mathcal{Q}_1$ can be rewritten as
    \begin{equation}\label{eq:pcrc}
    \begin{aligned}
    ||\bV\bW||_F^2=\frac{N_{\rm t}}{M_{\rm t}}||\bW||_F^2=||\mathbf{\mathbf{B}}||_F^2.
    \end{aligned}
    \end{equation}
Thus, we can reformulate problem $\mathcal{Q}_1$ as
    \begin{subequations}\label{eq:pcot}
    \begin{align}
    \mathcal{F}_1: \mathop{\mathrm{minimize}}\limits_{\mathbf{W},\mathbf{V}}&\ \ ||\mathbf{B}-\mathbf{VW}||_F^2\\
    {\rm s.t.}&\ \ \mathbf{V}\in \mathcal{S}_2,\\
    &\ \ ||\mathbf{W}||_F^2=||\mathbf{B}||_F^2\frac{M_{\rm t}}{N_{\rm t}}\triangleq\beta,
    \end{align}
    \end{subequations}
where $\mathcal{S}_2\in \left\{\mathcal{S}_{\rm CPS, 2}, \mathcal{S}_{\rm DPS, 2}\right\}$. Due to the phase shift constraint, problem $\mathcal{F}_1$ is still not easy to handle. We employ the variable projection and the inexact MM method \cite{gao2016energy} for this problem, which will be detailed as follows.

\subsection{Digital Precoder Design}

By assuming a fixed $\bV$, the optimization of the digital precoder $\bW$ can be formulated from problem $\mathcal{F}_1$ as follows
    \begin{subequations}\label{eq:pcow}
    \begin{align}
    \mathcal{F}_2: \mathop{\mathrm{minimize}}\limits_{\mathbf{W}}&\ \ ||\mathbf{B}-\mathbf{VW}||_F^2\\
    {\rm s.t.}&\ \ ||\bW||_F^2=\beta.
    \end{align}
    \end{subequations}
Considering the column orthogonality of matrix $\bV$, i.e., $\bV^H\bV=N_{\rm t}/M_{\rm t}\bI$, we can further simplify the above problem by first expanding the above objective function and get
    \begin{subequations}\label{eq:pcwe}
    \begin{align}
    \mathcal{F}_3: \mathop{\mathrm{minimize}}\limits_{\mathbf{W}}&\  \mathrm{Tr}\left\{\mathbf{B}^H\mathbf{B}-2\Re\left(\mathbf{B}^H\mathbf{VW}\right)+\bW^H\bV^H\bV\bW\right\}\\
    {\rm s.t.}&\  ||\bW||_F^2=\beta.
    \end{align}
    \end{subequations}
It can be seen that the last term of the objective function in problem $\mathcal{F}_3$ is a constant whose value is equal to $\beta N_{\rm t}/M_{\rm t}$. Then, after dropping the first and last terms of the objective function, problem $\mathcal{F}_3$ can be equivalently expressed as
    \begin{equation}\label{eq:pcwf}
    \begin{aligned}
    \mathcal{F}_4: \mathop{\mathrm{minimize}}\limits_{\mathbf{W}}&\ \ ||\bV^H\bB-\bW||_F^2\\
    {\rm s.t.}&\ \ ||\bW||_F^2=\beta,
    \end{aligned}
    \end{equation}
which is a projection problem, and its solution can be obtained in a closed-form, given by
    \begin{equation}\label{eq:pcws}
    \begin{aligned}
    \mathbf{W}=\sqrt{\beta}\frac{\bV^H\bB}{||\bV^H\bB||_F}.
    \end{aligned}
    \end{equation}

\subsection{Analog Precoder Design}
Let us now substitute formula \eqref{eq:pcws} back to the original problem $\mathcal{F}_1$, we have
    \begin{subequations}\label{eq:pcov}
    \begin{align}
    \mathcal{F}_5: \mathop{\mathrm{minimize}}\limits_{\mathbf{V}}&\ \ \Big|\Big|\bB-\sqrt{\beta}\bV\frac{\bV^H\bB}{\|\bV^H\bB\|_F}\Big|\Big|_F^2\\
    {\rm s.t.}&\ \ \bV \in \mathcal{S}_2,
    \end{align}
    \end{subequations}
whose objective function can be expanded as
    \begin{align}\label{eq:vofe}
    &\mathrm{Tr}\left\{\bB^H\bB-2\sqrt{\beta}\Re\left(\bB^H\bV\frac{\bV^H\bB}{||\bV^H\bB||_F}\right)\right\}\notag\\
    &\qquad \qquad +\beta\mathrm{Tr}\left\{\frac{\bB^H\bV}{||\bV^H\bB||_F}\bV^H\bV\frac{\bV^H\bB}{||\bV^H\bB||_F}\right\}\notag\\
    =&\mathrm{Tr}\left\{\bB^H\bB\right\}-2\sqrt{\beta}||\bV^H\bB||_F+\beta\frac{N_{\rm t}}{M_{\rm t}}.
    \end{align}
Note that the first and last terms of the right hand side in \eqref{eq:vofe} are constant. After ignoring them, problem $\mathcal{F}_5$ can be reformulated as
    \begin{subequations}\label{eq:ovff}
    \begin{align}
    \mathcal{F}_6: \mathop{\mathrm{minimize}}\limits_{\mathbf{V}}&\ \ -||\bV^H\bB||_F^2\\
    {\rm s.t.}&\ \ \bV \in \mathcal{S}_2,
    \end{align}
    \end{subequations}
which can be further converted into
    \begin{subequations}\label{eq:ovfc}
    \begin{align}
    \mathcal{F}_7: \mathop{\mathrm{minimize}}\limits_{\mathbf{V}}&\ \ -\sum_{i=1}^{M_{\rm t}}\bv_i^H\bC\bv_i\\
    {\rm s.t.}&\ \ \bV \in \mathcal{S}_2,
    \end{align}
    \end{subequations}
where $\bC=\bB\bB^H$ and $\bv_i$ is the $i$th column of the analog precoder $\bV$. For the convenience of derivation, let $\bD_{i}=\bC_{(i-1)\frac{N_{\rm t}}{M_{\rm t}}+1:i\frac{N_{\rm t}}{M_{\rm t}},(i-1)\frac{N_{\rm t}}{M_{\rm t}}+1:i\frac{N_{\rm t}}{M_{\rm t}}}$ and we can rewrite problem $\mathcal{F}_7$ as
    \begin{subequations}\label{eq:ovfd}
    \begin{align}
    \mathcal{F}_8: \mathop{\mathrm{minimize}}\limits_{\mathbf{V}}&\ \ -\sum_{i=1}^{M_{\rm t}}\bp_i^H\bD_i\bp_i\\
    {\rm s.t.}&\ \ \bV \in \mathcal{S}_2,
    \end{align}
    \end{subequations}
where $\bp_i$ is defined in \eqref{eq:pchv}.
Let $\br=[\bp_1^T,\bp_2^T,\ldots,\bp_{M_{\rm t}}^T]^T \in \mathbb{C}^{N_{\rm t}\times 1}$  and $\bA=\blkdiag{\bD_1,\ldots,\bD_{M_{\rm t}}}$. Then, problem $\mathcal{F}_8$ can be equivalently cast as
    \begin{subequations}\label{eq:ovfr}
    \begin{align}
    \mathcal{F}_9: \mathop{\mathrm{minimize}}\limits_{\mathbf{V}}&\ \ f(\br)=-\br^H\bA\br\label{eq:ovfra}\\
    {\rm s.t.}&\ \ \bV \in \mathcal{S}_2.\label{eq:ovfrb}
    \end{align}
    \end{subequations}

Note that the objective function of Problem $\mathcal{F}_9$ is quadratic convex and can be similarly transformed into
    \begin{align}\label{eq:ormm}
    \mathcal{F}_{10}: \mathop{\mathrm{minimize}}\limits_{\mathbf{V}\in \tilde{\mathcal{S}}_2}\ \ F_{\delta}(\mathbf{r})\triangleq f(\mathbf{r})-\delta||\mathbf{r}||_2^2,
    \end{align}
which can also be tackled by the inexact MM method with $\nabla_{\mathbf{r}}G_{\delta}(\mathbf{r}|\bar{\mathbf{r}})=-2\bA\br-2\delta \bar{\br}$. The whole procedure for problem $\mathcal{F}_1$ is shown in \textbf{Algorithm \ref{alg:algvmap}}.

\begin{algorithm}
\caption{\textbf{A}lternating \textbf{V}ariable \textbf{P}rojection and \textbf{I}nexact \textbf{MM} Based Hybrid \textbf{A}nalog/\textbf{D}igital \textbf{P}recoding (AVPIM-ADP)}
\label{alg:algvmap}
\begin{algorithmic}[1]
\REQUIRE The fully digital precoding matrix $\mathbf{B}$.
\STATE
Find analog precoder $\mathbf{V}$ by handling problem $\mathcal{F}_8$.
\STATE Digital precoder can be found through
\begin{align}
\mathbf{W}=\sqrt{\beta}\frac{\bV^H\bB}{||\bV^H\bB||_F}.\notag
\end{align}
\end{algorithmic}
\end{algorithm}

\subsection{Convergence and Complexity Analysis}
The convergence analysis of \textbf{Algorithm \ref{alg:algvmap}} is equivalent to that of the inexact MM method for deriving the analog precoder $\mathbf{V}$ in Step 3, and the interested readers may refer again to \cite{shao2019framework}. The computational complexity for deriving the digital precoder can be shown to be given by $\mathcal{O}(KN_{\rm t}M_{\rm t})$. For the analog precoder, the complexity in each iteration can be evaluated as $\mathcal{O}(N_{\rm t}^2)$. By assuming that the algorithm involves $J$ iterations, the whole complexity of \textbf{Algorithm \ref{alg:algvmap}} can be approximated by $\mathcal{O}(JN_{\rm t}^2+KN_{\rm t}M_{\rm t})$.

\section{Simulation Results}\label{sec:sim}
This section evaluates the performance of the proposed algorithms. The simulation setup is given in \tabref{tab:test}.
For all UTs, the channel space angles $\vartheta_k^{\mathrm{x}}$ and $\vartheta_k^{\mathrm{y}}$ are independent and follow the uniform distribution in the interval [-1,1). The adopted power consumption values are listed in \tabref{tab:pwct}.
The noise power $N_0$ is characterized by $N_0=k_{\rm B}B_{\rm w}T_{\rm n}$, where $k_{\rm B}$ and $T_{\rm n}$ denote the Boltzmann constant and the noise temperature, respectively.
The downlink channel power can be calculated by
\begin{align}\label{eq:chp}
\gamma_k = G_{\rm sat}G_{\rm ut}N_\mathrm{t}\left(\frac{v_{\rm c}}{4\pi f_{\rm c}d_k}\right)^2, \forall k,
\end{align}
where $G_{\rm sat}$ and $G_{\rm ut}$ are the antenna gains of the transmitter and the receiver, respectively. The carrier frequency is given by $f_{\rm c}$, $v_{\rm c}$ denote the speed of light and $d_k$ is the distance between the $k$th UT and the LEO satellite, which can be computed similarly as in \cite{li2020downlink}.

The inexact MM method for problem \eqref{eq:mmre} is set up as follows. The initial penalty parameter is set as $\eta = 0.01$. Once more than $J = 400$ MM iterations are executed or the distance between two successive objectives is less than $\epsilon = 10^{-5}$, it is multiplied by a factor of $c = 5$. When $\eta > 200$, \textbf{Algorithm \ref{alg:algiemm}} stops.
\begin{table}[!b]
\caption{Simulation Parameters}\label{tab:test}
\centering
\ra{1.3}
\footnotesize
\begin{tabular}{LR}
\toprule
Parameter & Value  \\
\midrule
\rowcolor{lightblue}
Number of satellite antennas $N_\mathrm{t}^{\mathrm{x}}\times N_\mathrm{t}^{\mathrm{y}}$     & $12\times 12$    \\
Number of UTs $K$ & 9\\
\rowcolor{lightblue}
Rician factor $\kappa_k$                & 18 dB  \\
Power amplifier efficiency $1/\xi$      &0.5                             \\
\rowcolor{lightblue}
System bandwidth $B_{\rm w}$ &20 MHz\\
Carrier frequency $f_c$ & 2 GHz\\
\rowcolor{lightblue}
Antenna gain $G_{\rm sat}$, $G_{\rm ut}$ & 3 dB, 0 dB\\
Speed of light $v_{\rm c}$ & 3$\times$ 10$^8$ m/s \\
\rowcolor{lightblue}
Boltzmann constant $k_{\rm B}$ &$1.38\times 10^{-23}$ J$\cdot\text{K}^{-1}$\\
Noise temperature $T_{\rm n}$ & $300$ K\\
\bottomrule
\end{tabular}
\end{table}

\begin{table}[!t]
\caption{Typical Power Consumption Values \cite{mendezrial2016hybrid}}\label{tab:pwct}
\centering
\ra{1.3}
\footnotesize
\begin{tabular}{LR}
\toprule
Parameter & Value (mW)  \\
\midrule
\rowcolor{lightblue}
$P_{\mathrm{DAC}}=P_{\mathrm{ADC}}$              & 300                         \\
$P_{\mathrm{mixer}}$                     & 19                      \\
\rowcolor{lightblue}
$P_{\mathrm{LPF}}$                & 14  \\
$P_{\mathrm{BBA}}$                    & 5                       \\
\rowcolor{lightblue}
$P_{\mathrm{LO}}$              & 5                   \\
$P_{\mathrm{BB}}$   & 200               \\
\rowcolor{lightblue}
$P_{\mathrm{Ps}}(\infty), P_{\mathrm{Ps}}(4), P_{\mathrm{Ps}}(3), P_{\mathrm{Ps}}(2)$                         &  25, 20, 16, 12          \\
$P_{\mathrm{syn}}$ &50 \\
\rowcolor{lightblue}
$P_{\mathrm{LNA}}$ &20 \\
\bottomrule
\end{tabular}
\end{table}

\begin{figure}[!b]
\centering
\subfloat[Fully connected architecture.]{\includegraphics[width=0.45\textwidth]{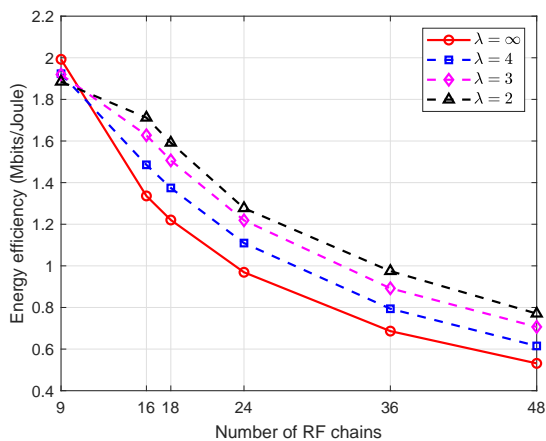}\label{EE_Mt_resolution_f}}
\hfill
\subfloat[Partially connected architecture.]{\includegraphics[width=0.45\textwidth]{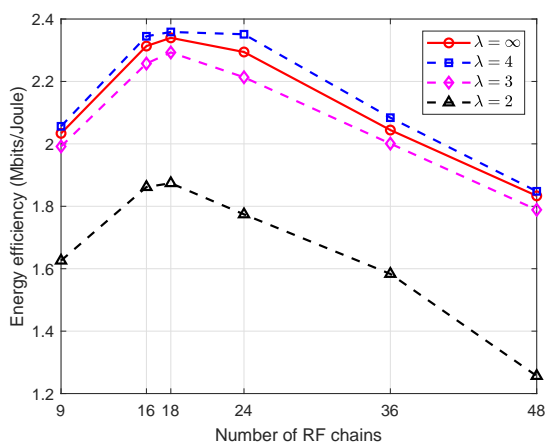}\label{EE_Mt_resolution_p}}
\caption{EE versus the number of RF chains with $P=10$ dBW and different resolutions $\lambda$.}
\label{EE_Mt_resolution}
\end{figure}

    \begin{figure}[!t]
		\centering
		\includegraphics[width=0.45\textwidth]{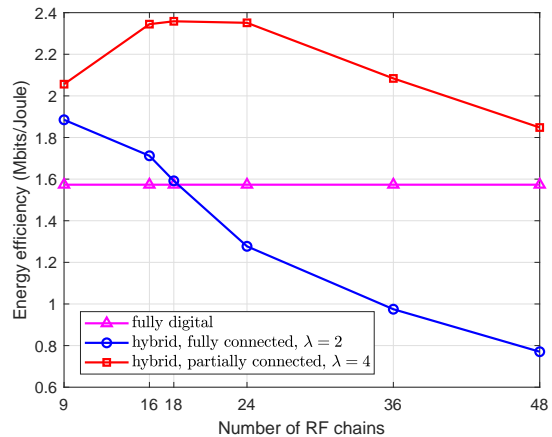}
		\caption{EE versus the number of RF chains with $P=10$ dBW, for different architectures.}
        \label{EE_Mt_structure}
	\end{figure}

First, we compare the hybrid precoding designs considering the phase shifters with both finite and infinite resolutions. Fig. \ref{EE_Mt_resolution} evaluates the relationship between the EE and the number of RF chains under both the fully and partially connected architectures.
We can observe that for the fully connected architecture, when the number of RF chains is small, higher EE is achieved.
Besides, as the resolution increases, the EE performance becomes worse in general.
The explanation is straightforward.
According to formula \eqref{eq:fptd}, as the resolution of phase shifters increases a little, the power consumption is significantly increased while the rate presents a small improvement, leading to a decrease in the value of EE.
However, with less number of RF chains, the rate dominates the performance of EE and thus, the infinite resolution scenario leads to better EE performance.
In addition, for the partially connected architecture, the EE performance first increases with an increase in the number of the employed RF chains.  Though, after some point, there is a decrease in the value of EE, since the EE performance is dominated by the static power consumption due to the large number of RF chains.
Moreover, the case with the resolution of the phase shifters $\lambda=4$ presents the best EE performance due to the significant rate performance at relatively lower power consumption.

\begin{figure}[!t]
\centering
\subfloat[Upper bound of EE.]{\includegraphics[width=0.45\textwidth]{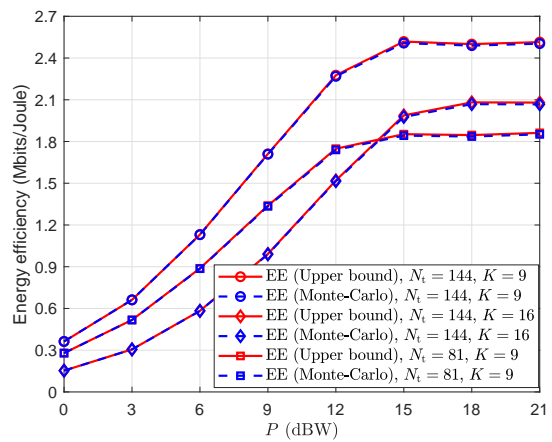}\label{EEUpperBound}}
\hfill
\subfloat[Upper bound of the sum rate.] {\includegraphics[width=0.45\textwidth]{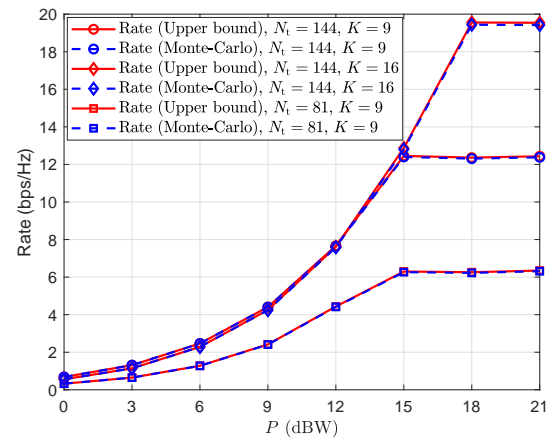}\label{RateUpperBound}}
\caption{The achievable sum rate and EE versus power budget with different numbers of antennas $N_{\rm t}$ and UTs $K$.}
\label{Upperbound}
\end{figure}

\begin{figure}[htbp]
\centering
\subfloat[Fully connected architecture.]{\includegraphics[width=0.45\textwidth]{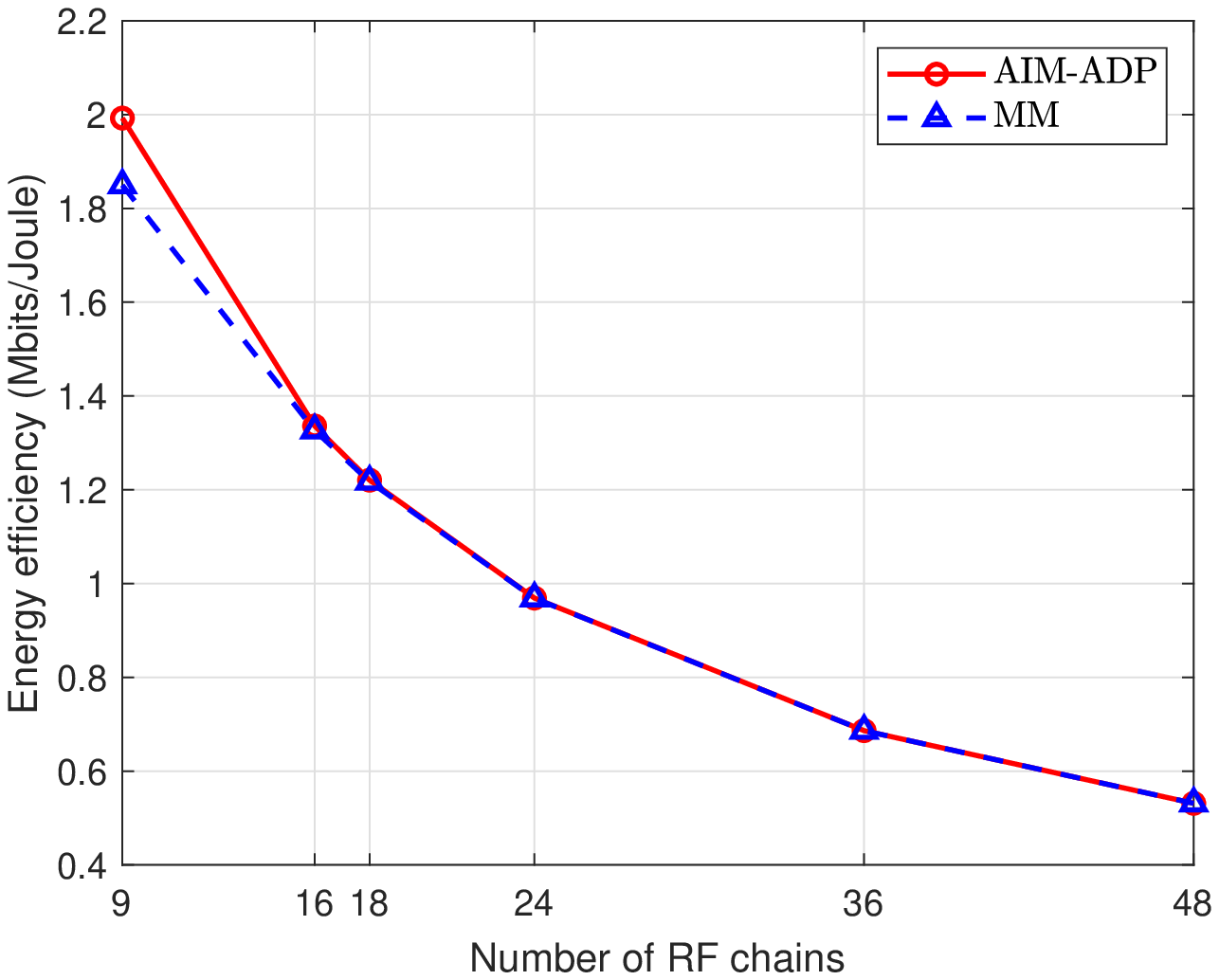}\label{EE_Mt_method_co_f}}
\hfill
\subfloat[Partially connected architecture.]{\includegraphics[width=0.45\textwidth]{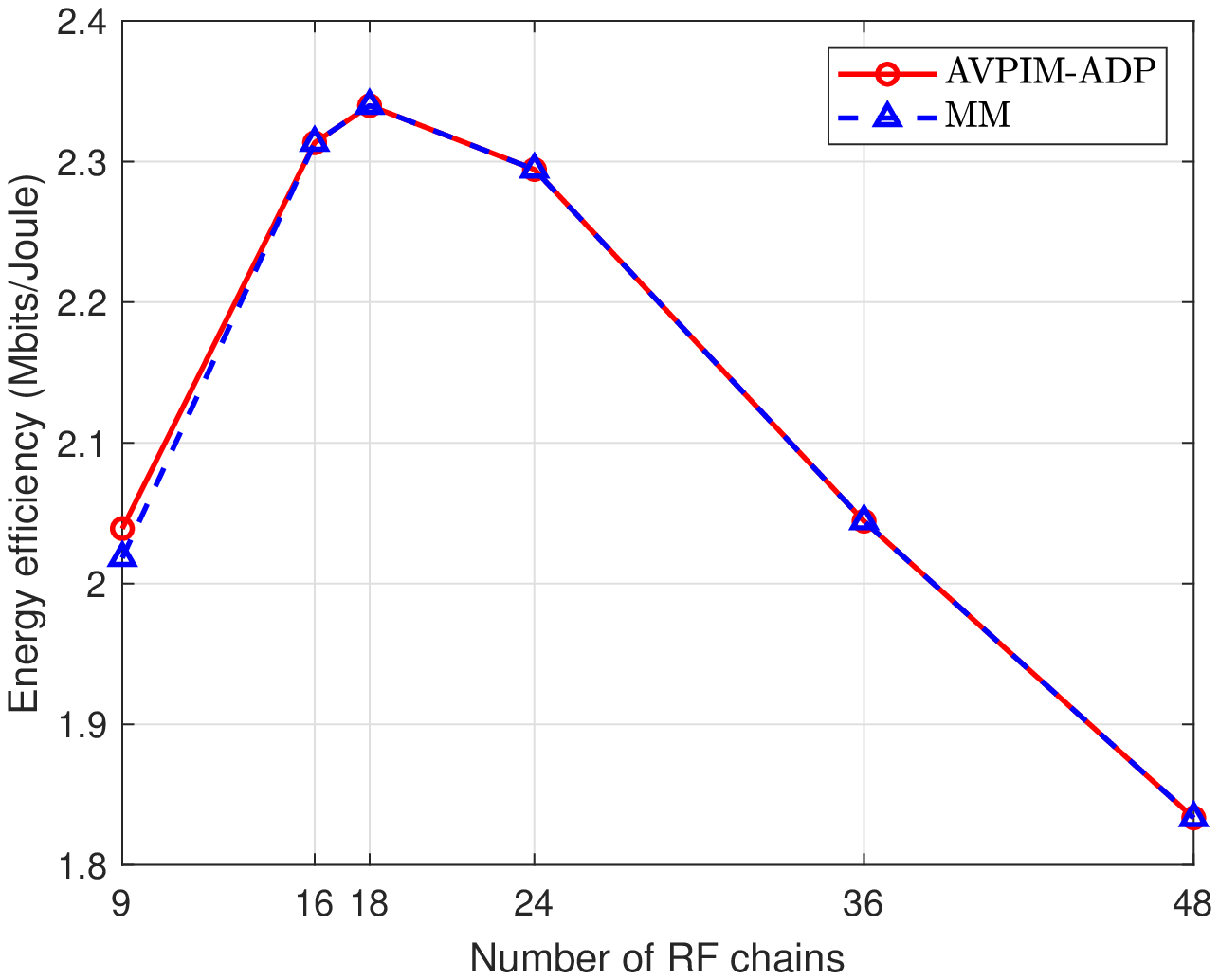}\label{EE_Mt_method_co_p}}
\caption{EE versus the number of RF chains with $P=10$ dBW, $\lambda=\infty$ and different solutions for the analog precoder $\bV$.}
\label{EE_Mt_method_co}
\end{figure}

\begin{figure}[htbp]
\centering
\subfloat[Fully connected architecture.]{\includegraphics[width=0.45\textwidth]{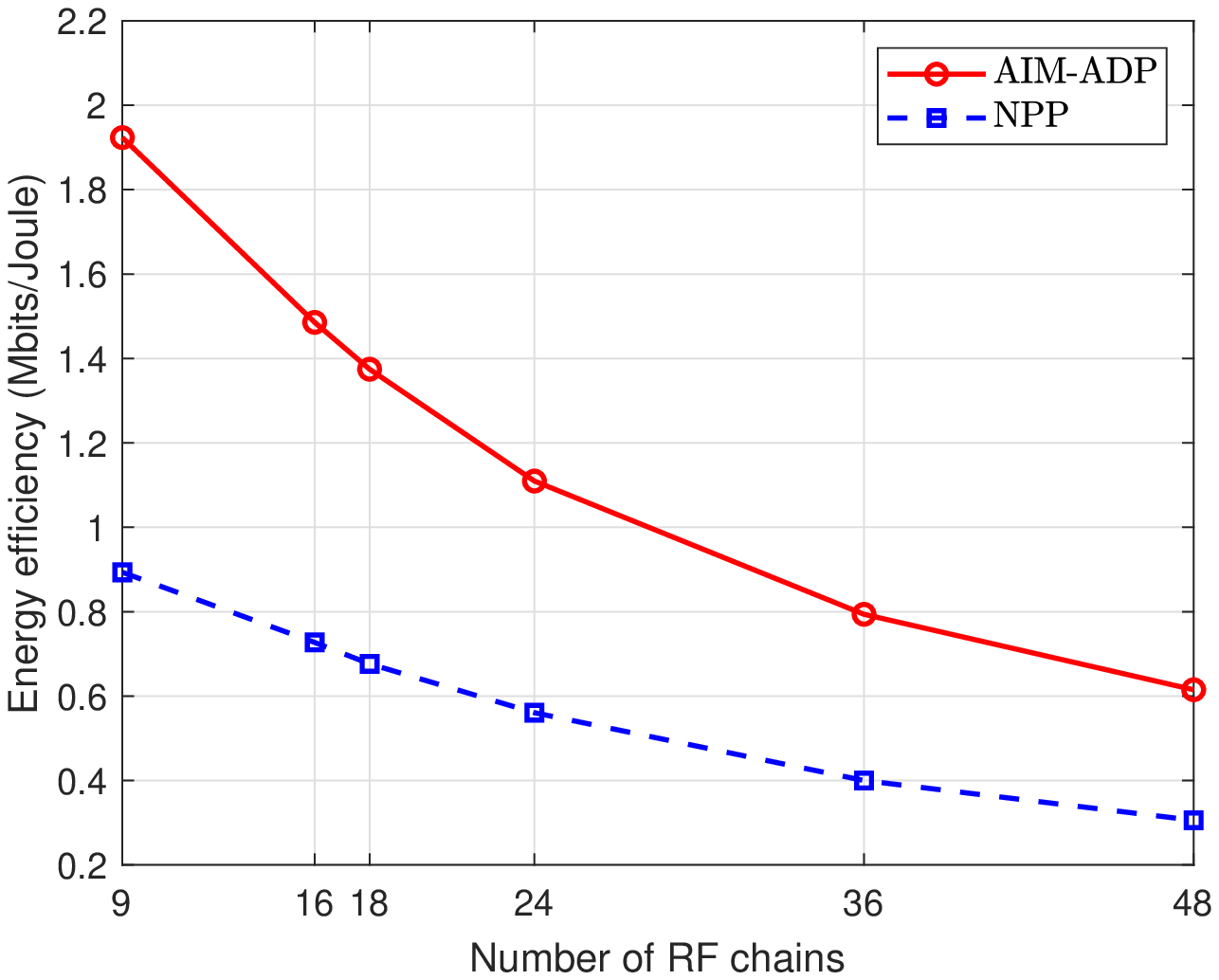}\label{EE_Mt_method_leq4_f}}
\hfill
\subfloat[Partially connected architecture.]{\includegraphics[width=0.45\textwidth]{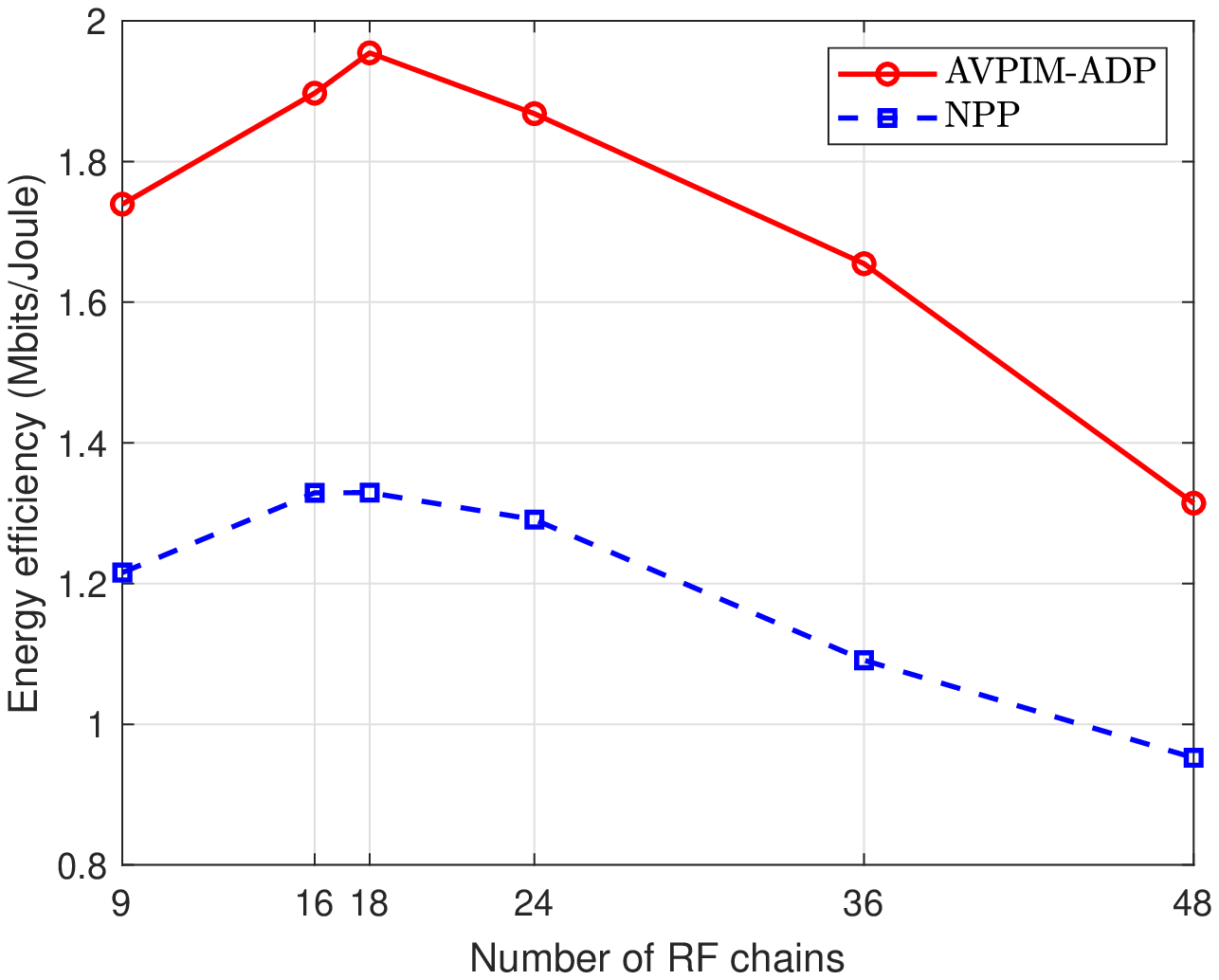}\label{EE_Mt_method_leq4_p}}
\caption{EE versus the number of RF chains with $P=10$ dBW, $\lambda=4$ and different solutions for the analog precoder $\bV$.}
\label{EE_Mt_method_leq4}
\end{figure}

In Fig. \ref{EE_Mt_structure}, the EE performance of the proposed hybrid architectures is compared to that of the fully digital system that serves as a benchmark. It can be seen that it is better to apply the partially connected architecture, which is a more energy efficient solution under the given scenario. Meanwhile, the EE performance of the fully digital and the fully connected architectures are generally worse. This can be explained through the comparison of formulae \eqref{eq:fptd}, \eqref{eq:pptd}, and \eqref{eq:fdpt}. By inspecting the aforementioned equations, it can be deduced that the power consumed by the fully connected architecture is nearly $M_{\rm t}$ times of that consumed by the partially connected one.
Hence, it is suitable only for systems with a relatively small number of RF chains, as observed.
In addition, the very high power consumption of the fully digital architecture, dominates its EE, resulting in the observed worst performance among the three solutions, in general. Therefore, with a hybrid architecture, we can achieve lower cost and hardware complexity with better EE by employing phase shifters of relatively lower resolutions and fewer RF chains.
Besides, the power consumption of hybrid precoding increases linearly with the increasing of the number of RF chains while the achievable rate increases much more slowly, which results in a gradually decreasing slope of the corresponding EE curve as the number of RF chains increases.


Fig. \ref{Upperbound} illustrates the EE performance versus the power budget under different scenarios. In general, the EE first rises to a peak and then nearly remains steady as the power budget increases. The reason can be explained as follows. In particular, there exists a threshold of the transmit power which maximizes the EE. After the given power budget achieves that, the actual value of the transmit power actually remains constant. Therefore, when the power budget increases to that point, the corresponding EE also reaches its maximum value and then becomes saturated. In addition, Fig. \ref{Upperbound} demonstrates the tightness of the adopted upper bound in \eqref{eq:exrt} in terms of the sum rate and EE with different values of the key parameters, e.g., the number of antennas and the UTs.

In Figs. \ref{EE_Mt_method_co}--\ref{EE_Mt_method_leq4}, we compare the performance of the proposed approach to the state-of-the-art methods in \cite{arora2019hybrid} and \cite{guo2020weighted} under the fully or the partially connected architectures. In the former, the MM method is applied for phase shifters of infinite resolution. In the latter, the NPP method is adopted for phase shifters of infinite resolution. For both the infinite and finite resolution cases, the two applied methods, namely AIM-ADP and AVPIM-ADP, present slight performance gains over the existing ones, namely MM and NPP.

\section{Conclusion}\label{sec:conc}
This paper investigated the EE maximization based hybrid precoding design, where the analog precoder is implemented with either continuous or discrete phase shift network, for the downlink massive MIMO LEO satellite transmission considering both the fully and the partially connected architectures. To tackle this fractional programming problem under the nonconvex phase constraints, we divided it into two subproblems. First, the precoding vectors for the fully digital system were obtained by developing an algorithmic solution based on Dinkelbach's algorithm and the WMMSE method. Then, the Euclidean distance between the fully digital precoder and the analog/digital precoders was minimized to handle the latter subproblem via utilizing the alternating minimization approach and an inexact MM algorithm. Numerical results demonstrated the performance gains of the proposed algorithms over the existing approaches, especially for the case of finite resolution phase shifters.
\begin{appendices}
\section{Proof for the Upper bound in \eqref{eq:exrt}}\label{app:a}
We introduce an auxiliary function $f(x)$, whose expression is given by
    \begin{align}
    f(x)&=\log\left( \frac{ax}{bx+c}+1\right),\ x\geq 0, a,b,c\geq 0.
    \end{align}
It is not difficult to calculate the second derivative of the function $f(x)$, given by
\begin{align}
f''(x)=-\frac{ac\left((a+b)(bx+c)+((a+b)x+c)b\right)}{\left((a+b)x+c\right)^2\left(bx+c\right)^2}\leq 0.
\end{align}
Therefore, it can be decided that $f(x)$ is concave with non-negative parameters $a$, $b$, $c$ and $x\geq 0$ \cite{boyd2004convex}. Based on the above conclusions, let $a=|\mathbf{v}_k^H\mathbf{b}_k|^2, b=\sum_{{\ell}\neq k}|\mathbf{v}_k^H\mathbf{b}_{\ell}|^2, c= N_0$ and $x=\abs{g_k}^2$, which are all obviously non-negative. Hence, we can obtain that the logarithmic expression inside the expectation operator of the rate function $R_k$ is concave with respect to $\abs{g_k}^2$. Subsequently, by utilizing Jensen's inequality, i.e., $\mathbb{E}\left\{f\left(X\right)\right\}\leq f\left(\mathbb{E}\left\{X\right\}\right)$, we obtain
    \begin{equation}
    \begin{aligned}
    R_k&=\mathbb{E}\left\{\log\left(1+\frac{|\mathbf{b}_k^H\mathbf{h}_k|^2}{\sum_{\ell\neq k}|\mathbf{b}_{\ell}^H\mathbf{h}_k|^2+N_0}\right)\right\}\\
    &\leq \bar{R}_k
    \triangleq\log\left(1+\frac{|\mathbf{b}_k^H\mathbf{v}_k|^2\mathbb{E}\left\{|g_k|^2\right\}}{\sum_{{\ell}\neq k}|\mathbf{b}_{\ell}^H\mathbf{v}_k|^2\mathbb{E}\left\{|g_k|^2\right\}+N_0}\right)\\
    &\qquad \ \ =\log\left(1+\frac{\gamma_k|\mathbf{v}_k^H\mathbf{b}_k|^2}{\sum_{{\ell}\neq k}\gamma_k|\mathbf{v}_k^H\mathbf{b}_{\ell}|^2+N_0}\right).
    \end{aligned}
    \end{equation}
\end{appendices}


\end{document}